\documentclass[aps, prd, showkeys, showpacs ]{revtex4}

\usepackage{amsmath,amsfonts,amscd,amssymb,epsf, multirow}
\usepackage[small]{caption}

\newcommand{\pa}{\partial}

\newcommand{\vep}{\varepsilon}

\begin{document}

\title{  Casimir interaction  between spheres  in $\boldsymbol{(D+1)}$-dimensional Minkowski spacetime}

\author{L. P. Teo}
 \email{LeePeng.Teo@nottingham.edu.my}
 \affiliation{Department of Applied Mathematics, Faculty of Engineering, University of Nottingham Malaysia Campus, Jalan Broga, 43500, Semenyih, Selangor Darul Ehsan, Malaysia.}
\begin{abstract}
We consider the Casimir interaction between two spheres in $(D+1)$-dimensional Minkowski spacetime due to the vacuum fluctuations of scalar fields. We consider combinations of Dirichlet and Neumann boundary conditions. The TGTG formula of the Casimir interaction energy is derived. The computations of the T matrices of the two spheres are straightforward. To compute the two G matrices, known as translation matrices, which relate the hyper-spherical waves in two spherical coordinate frames differ by a translation, we generalize the operator approach employed in [IEEE Trans. Antennas Propag. \textbf{36}, 1078 (1988)]. The result is expressed in terms of an integral over Gegenbauer polynomials. In contrast to the $D=3$ case, we do not re-express the integral in terms of $3j$-symbols and hyper-spherical waves, which in principle, can be done but does not simplify the formula. Using our expression for the Casimir interaction energy, we derive the large separation and small separation asymptotic expansions of the Casimir interaction energy. In the large separation regime, we find that the Casimir interaction energy is of order $L^{-2D+3}$, $L^{-2D+1}$ and $L^{-2D-1}$ respectively for Dirichlet-Dirichlet, Dirichlet-Neumann and Neumann-Neumann boundary conditions, where $L$ is the center-to-center distance of the two spheres. In the small separation regime, we confirm that the leading term of the Casimir interaction agrees with the proximity force approximation, which is of order $d^{-\frac{D+1}{2}}$, where $d$ is the distance between the two spheres. Another main result of this work is the analytic computations of the next-to-leading order term in the small separation asymptotic expansion. This term is computed using careful order analysis as well as perturbation method. In the case the radius of one of the sphere goes to infinity, we find that the results agree with the one we derive for sphere-plate configuration. When $D=3$, we also recover previously known results. We find that when $D$ is large, the ratio of the next-to-leading order term to the leading order term is linear in $D$, indicating a larger correction at higher dimensions. The methodologies employed in this work and the results obtained can be used to study the one-loop effective action of the system of two spherical objects in the universe.
\end{abstract}
\pacs{03.70.+k, 11.10.Kk}
\keywords{  Casimir interaction, sphere-sphere configuration, higher dimensional spacetime, scalar field, analytic correction to proximity force approximation, large separation behavior}

\maketitle
\section{Introduction}
Motivated by the advent of string theory and the endeavor to solve fundamental problems in physics such as dark energy and cosmological constant problem,   studying  physics in higher dimensional spacetime has become a norm rather than exception.

Casimir effect \cite{8} which was proposed more than 60 years ago  plays an important role in high energy physics since it is intimately related to the one-loop effective action of a quantum field \cite{9}. Although most of the works in Casimir effect were focused in (3+1)-dimensional spacetime, there have been quite a considerable amount of work on Casimir effect in higher dimensional spacetime. One of the pioneering works is the work \cite{10} where the Casimir effect of a $D$-dimensional rectangular cavity is studied. Subsequently, Casimir effect of a $D$-dimensional spherical cavity were also considered \cite{11,12,13}. Nonetheless, in contrast  to the Casimir interaction between two parallel plates considered by Casimir in his pioneering work \cite{8}, the Casimir energy of a $D$-dimensional rectangular or spherical cavity is  a self-energy rather than interaction energy.

In the end of last century, it has gradually been realized that the Casimir interaction between two objects should play a more important role since this is physically observable. Partly also due to the need to compare to Casimir experiments, researchers have started to research on the method to compute the Casimir interactions between two objects, in particular between a sphere and a plate. In the early phase of this research, most of the methods proposed were numerical \cite{14, 15, 16, 17, 18, 19, 20}. About eight years ago, new light has been shed on this problem by two new developments. First, Gies  and collaborators \cite{21, 22, 23, 24, 25}   used worldline representation to compute the Casimir interaction between two objects imposed with Dirichlet boundary conditions. Later  a few groups of researchers \cite{55,26,27,28,43,6,44,45,32,33,34,36,37,38,39}   independently developed some methods to compute   functional representations for the Casimir interaction between specific pairs of objects which  all used the idea of  multiple scattering in some form. In a time span of three years, a scheme has been worked out which in principle can be used to compute the Casimir interactions between any two objects \cite{6,32}. Motivated by the mode summation approach proposed for the cylinder-cylinder configuration \cite{40, 41, 42}, we managed to interpret the scheme proposed in \cite{6,32} from the point of view of mode summation approach \cite{7}. Despite that we only considered examples in $(3+1)$-dimensional spacetime, it is quite easy to see that the mathematical scheme we deployed in \cite{7} is not limited to $(3+1)$-dimensional spacetime.

There have been quite a number of works on the Casimir effect between two parallel plates in higher dimensional spacetime. The case of $(D+1)$-dimensional Minkowski spacetime has been treated in \cite{10} as limiting case of rectangular cavity. As a first step to consider Casimir interaction between nontrivial objects in higher dimensional spacetime, we have considered the Casimir interaction between a sphere and a plate in $(D+1)$-dimensional Minkowski spacetime in \cite{1}. In this work, we consider two spheres with Dirichlet or Neumann boundary conditions. This scenario is more interesting since it can be used to model two spherical objects in the universe. The mathematical scheme we developed in \cite{7} is used to compute the Casimir interaction energy. In  hyper-spherical coordinates, it is easy to write down the scattering matrices of the two spheres. The most technical part of the problem is to derive the translation matrices which relate the hyper-spherical coordinate system centered at the two spheres. For this, we generalize the operator approach developed for the $D=3$ case \cite{31, 7}. A major difference is that we do not use $3j$-symbols to rewrite the translation matrices as linear combinations of spherical waves but leave it as integrals over Gegenbauer polynomials.

After deriving the functional representation for the Casimir interaction energy, a natural question to ask is what we can infer from the formula. One of the most important things we want to know is the strength of the interaction. To have an idea of this, we need to compute the asymptotic expansions of the Casimir interaction energy at small and large separations.  The computation of the large separation asymptotic behavior is quite straightforward since it only involves a few terms which can be computed explicitly. For the small separation asymptotic expansion, the computations are more complicated and a careful order counting is needed. For $D=3$ case, such analysis have been carried out in \cite{33,34,46,47,48,2,50,51,5} for different configurations. Our current scenario is closest to \cite{5} where two spheres in $(3+1)$-dimensional spacetime is considered. However, a major difference is that we do not have $3j$-symbols in the formula for the Casimir interaction energy. So instead of the integral representation for the $3j$-symbols, we have to use an integral representation for the Gegenbauer polynomials. It turns out that this is in fact not any more complicated.

This work will shed some light on how to compute the Casimir effect between two objects with nontrivial geometry in higher dimensional spacetime. It will also be interesting for researchers wanted to study quantum system of two spherical objects in higher dimensional spacetime.

\section{The Casimir interaction between a sphere and a plate}
In this section, we consider the Casimir interaction energy between two spheres in $(D+1)$-dimensional Minkowski spacetime equipped with the standard metric
$$ds^2=dt^2-dx_1^2-\ldots-dx_D^2.$$
The equation of motion of a scalar field $\varphi$ is
\begin{equation}\label{eq12_3_2}
\left(\frac{\pa^2}{\pa x_1^2}+\ldots+\frac{\pa^2}{\pa x_D^2}\right)\varphi = -\frac{\omega^2}{c^2}\varphi.
\end{equation}

 Assume that the   radii of the spheres are $R_1$ and $R_2$ respectively, and the centers are at $O_1=(0, 0, \ldots, 0)$ and   $O_2=(L, 0, \ldots, 0)$ respectively.
We use the hyper-spherical coordinate system: \begin{align*}
x_1=&r\cos\theta_1\\
x_2=&r\sin\theta_1\cos\theta_2\\
&\vdots\\
x_{D-1}=&r\sin\theta_1\ldots\sin\theta_{D-2}\cos\theta_{D-1}\\
x_D=&r\sin\theta_1\ldots\sin\theta_{D-2}\sin\theta_{D-1}
\end{align*}
When $\boldsymbol{r}=(x_1,\ldots,x_D)$ ranges over $\mathbb{R}^D$, $r$ ranges from 0 to $\infty$, whereas
\begin{align*}
0\leq \theta_i\leq\pi, \quad i=1,2,\ldots,D-2,
\end{align*} and
$$-\pi\leq\theta_{D-1}\leq\pi.$$In the following, we will denote by $S^{D-1}$ the region
\begin{align*}
0\leq \theta_i\leq\pi, \quad i=1,2,\ldots,D-2;\quad -\pi\leq\theta_{D-1}\leq\pi.
\end{align*}
The volume element $\displaystyle\prod_{i=1}^D dx_i$ is equal to $$r^{D-1}\prod_{i=1}^{D-1}\sin^{D-i-1}\theta_id\theta_i $$ in spherical coordinates. Denote by $d\Omega_{D-1}$ the measure
$$\prod_{i=1}^{D-1}\sin^{D-i-1}\theta_id\theta_i.$$ Then
$$\int_{S^{D-1}}d\Omega_{D-1}=\frac{2\pi^{\frac{D}{2}}}{\Gamma\left(\frac{D}{2}\right)}$$ is the volume of the unit sphere $x_1^2+x_2^2+\ldots+x_{D}^2=1$.

In spherical coordinates, the equation of motion \eqref{eq12_3_2}
becomes
\begin{equation*}
\left(\frac{\pa^2}{\pa r^2}+\frac{D-1}{r}\frac{\pa}{\pa r}+\frac{1}{r^2}\sum_{i=1}^{D-1}\frac{1}{\prod_{j=1}^{i-1}\sin^2\theta_j}\left(\frac{\pa^2}{\pa\theta_i^2}
+(D-i-1)\frac{\cos\theta_i}{\sin\theta_i}\frac{\pa}{\pa\theta_i}\right)\right)\varphi=-\frac{\omega^2}{c^2}\varphi.
\end{equation*}
The solutions of this differential equation are parametrized by $\boldsymbol{m}=(m_1,\ldots,m_{D-1})$, with
$$l=m_1\geq m_2\geq \ldots\geq m_{D-2}\geq |m_{D-1}|.$$The regular and outgoing spherical waves are \cite{29,30}:
\begin{equation}\label{eq12_3_6}
\begin{split}
\varphi_{\boldsymbol{m}}^{\text{reg}}(\boldsymbol{x},k)=\mathcal{C}_l^{\text{reg}}C_{\boldsymbol{m}}j_l( k r)\boldsymbol{Y}_{\boldsymbol{m}}(\boldsymbol{\theta}),\\
\varphi_{\boldsymbol{m}}^{\text{out}}(\boldsymbol{x},k)=\mathcal{C}_l^{\text{out}}C_{\boldsymbol{m}}h_l^{(1)}( k r)\boldsymbol{Y}_{\boldsymbol{m}}(\boldsymbol{\theta}),
\end{split}
\end{equation}
where
\begin{align*}
j_l(z)=z^{-\frac{D-2}{2}}J_{l+\frac{D-2}{2}}(z),\hspace{1cm}h_l^{(1)}(z)=z^{-\frac{D-2}{2}}H^{(1)}_{l+\frac{D-2}{2}}(z),
\end{align*}
\begin{align*}
\boldsymbol{Y}_{\boldsymbol{m}}(\boldsymbol{\theta})=e^{im_{D-1}\theta_{D-1}}\prod_{j=1}^{D-2}\sin^{|m_{j+1}|}\theta_{j}C_{m_j-|m_{j+1}|}^{|m_{j+1}|+\frac{D-j-1}{2}}(\cos\theta_j),
\end{align*}$J_{l+\frac{D-2}{2}}(z)$ and $H^{(1)}_{l+\frac{D-2}{2}}(z)$ are Bessel functions, and $C_n^{\nu}(z)$  is a Gegenbauer polynomial  defined by
\begin{align*}
(1-2zt+t^2)^{-\nu}=\sum_{n=0}^{\infty} C_n^{\nu}(z)t^n.
\end{align*} The Gegenbauer polynomials satisfy the orthogonality relation
\begin{align}\label{eq12_3_11}
\int_{-1}^1dx C_n^{\nu}(x)C_m^{\nu}(x)(1-x^2)^{\nu-\frac{1}{2}}=\frac{\pi 2^{1-2\nu}\Gamma(n+2\nu)}{n!(n+\nu)\Gamma(\nu)^2}\delta_{n,m}.
\end{align}
Hence, the hyperspherical harmonics $\boldsymbol{Y}_{\boldsymbol{m}}(\boldsymbol{\theta})$ satisfy the orthogonality condition
\begin{equation*}
\int_{S^{D-1}} d\Omega_{D-1}\boldsymbol{Y}_{\boldsymbol{m}}(\boldsymbol{\theta}) \boldsymbol{Y}_{\boldsymbol{m}'}(\boldsymbol{\theta})^* =\frac{1}{C_{\boldsymbol{m}}^2}\delta_{\boldsymbol{m},\boldsymbol{m}'},
\end{equation*}
where
\begin{align*}
C_{\boldsymbol{m}}=\sqrt{\frac{1}{2\pi^{D-1}}\prod_{j=1}^{D-2}\frac{2^{2|m_{j+1}|+D-j-2}\Gamma\left(|m_{j+1}|+\frac{D-j-1}{2}\right)^2\left(m_j+\frac{D-j-1}{2}\right)(m_j-|m_{j+1}|)!}
{\Gamma\left(m_j+|m_{j+1}|+D-j-1\right)}}.
\end{align*}
The constants $\mathcal{C}_l^{\text{reg}}$ and $\mathcal{C}_l^{\text{out}}$ are defined by
\begin{align*}
\mathcal{C}_l^{\text{reg}}=i^{-l},\quad \mathcal{C}_l^{\text{out}}=\frac{\pi}{2}i^{l+D-1},
\end{align*}  so that
\begin{align*}
\mathcal{C}_l^{\text{reg}}j_l(iz)=z^{-\frac{D-2}{2}}I_{l+\frac{D-2}{2}}(z),\hspace{1cm}\mathcal{C}_l^{\text{out}}h^{(1)}_l(iz)=z^{-\frac{D-2}{2}}K_{l+\frac{D-2}{2}}(z).
\end{align*}

In \cite{7}, we have discussed the mathematical formalism underlying the TGTG formula for the Casimir interaction energy
\begin{equation}\label{eq12_3_1}
E_{\text{Cas}}=\frac{\hbar}{2\pi}\int_0^{\infty} d\xi \ln\det\left(1-\mathbb{T}^1\mathbb{G}^{1}\mathbb{T}^2\mathbb{G}^{2}\right),
\end{equation}  between two objects. It is easy to see that this formalism does not depend on the dimension of spacetime and the type of quantum field involved. It can be applied for Casimir interaction in $(D+1)$-dimensional spacetime.
The $\mathbb{T}^1$ and $\mathbb{T}^2$ in this formula  are the Lippmann-Schwinger T-operators of the two spheres, which are related to the scattering matrices of the spheres. As in \cite{7}, it is easy to find that for Dirichlet (D) and Neumann  (N) boundary conditions, they are diagonal in $\boldsymbol{m}$   with diagonal elements given   by
\begin{equation*}
\begin{split}
&T_{\boldsymbol{m}}^{i, \text{D}}(\kappa)=T_{l}^{i, \text{D}}(\kappa)=\frac{I_{l+\frac{D-2}{2}}(\kappa R_i)}{K_{l+\frac{D-2}{2}}(\kappa  R_i)},\\
&T_{\boldsymbol{m}}^{i, \text{N}}(\kappa)=T_{l}^{i, \text{N}}(\kappa)=\frac{-\frac{D-2}{2}I_{l+\frac{D-2}{2}}(\kappa R_i)+\kappa R_iI_{l+\frac{D-2}{2}}'(\kappa R_i)}
{-\frac{D-2}{2}K_{l+\frac{D-2}{2}}(\kappa R_i)+\kappa R_i K_{l+\frac{D-2}{2}}'(\kappa R_i)}.
\end{split}
\end{equation*}Here $\kappa=\xi/c$ and $k=i\kappa$.

The translation matrices $\mathbb{G}^{1} $ and $\mathbb{G}^{2}$ in \eqref{eq12_3_1} are defined by
\begin{equation}\label{eq12_3_5}
\varphi_{\boldsymbol{m}'}^{\text{out}}(\boldsymbol{x}-\boldsymbol{L},k)=\sum_{\boldsymbol{m}}G^1_{\boldsymbol{m},\boldsymbol{m}'}\varphi_{\boldsymbol{m}}^{\text{reg}}(\boldsymbol{x},k),\end{equation}
\begin{equation}\label{eq3_4_1}
\varphi_{\boldsymbol{m}}^{\text{out}}(\boldsymbol{x}'+\boldsymbol{L},k)=\sum_{\boldsymbol{m}'}G^2_{\boldsymbol{m}',\boldsymbol{m}}\varphi_{\boldsymbol{m}'}^{\text{reg}}(\boldsymbol{x}',k),
\end{equation}where the summation over $\boldsymbol{m}$ is
\begin{align*}
\sum_{\boldsymbol{m}}=\sum_{l=0}^{\infty}\sum_{m_2=0}^{l}\sum_{m_3=0}^{m_2}\ldots\sum_{m_{D-2}=0}^{m_{D-3}}\sum_{m_{D-1}=-m_{D-2}}^{m_{D-2}}.
\end{align*}In the following, we will derive the explicit expressions for $G^1_{\boldsymbol{m},\boldsymbol{m}'}$ and $G^2_{\boldsymbol{m}',\boldsymbol{m}}$.

Express $\boldsymbol{k}=(k_1,k_2,\ldots,k_D)$ in hyper-spherical coordinates:
\begin{equation}\label{eq12_3_8}\begin{split}
k_1=&k\cos\theta^k_1,\\
k_2=&k\sin\theta^k_1\cos\theta^k_2,\\
&\vdots\\
k_{D-1}=&k\sin\theta^k_1\ldots\sin\theta^k_{D-2}\cos\theta^k_{D-1},\\
k_{D}=&k\sin\theta^k_1\ldots\sin\theta^k_{D-2}\sin\theta^k_{D-1},\end{split}
\end{equation}
and let $S^{D-1}_k$ be the region
\begin{gather*}
0\leq \theta^k_j\leq \pi,\quad 1\leq j\leq D-2,\quad
-\pi\leq \theta^k_{D-1}\leq\pi.
\end{gather*}
As is shown in \cite{30},
\begin{align}\label{eq12_3_4}
\int_{S^{D-1}_k}d\Omega_{D-1}^k\boldsymbol{Y}_{\boldsymbol{m}}(\boldsymbol{\theta}_k)e^{i\boldsymbol{k}\cdot\boldsymbol{r}}=(2\pi)^{\frac{D}{2}} i^{-l}j_l(kr)\boldsymbol{Y}_{\boldsymbol{m}}(\boldsymbol{\theta}).
\end{align}
This is equivalent to
\begin{align*}
e^{i\boldsymbol{k}\cdot\boldsymbol{r}}= (2\pi)^{\frac{D}{2}} \sum_{\boldsymbol{m}}
i^{-l} C_{\boldsymbol{m}}^2j_l(kr)\boldsymbol{Y}_{\boldsymbol{m}}(\boldsymbol{\theta})\boldsymbol{Y}_{\boldsymbol{m}}(\boldsymbol{\theta}_k)^*.
\end{align*}Notice that
\begin{align*}
j_0(kr)=a_D\int_{S_k^{D-1}} d\Omega_{D-1}^ke^{i\boldsymbol{k}\cdot\boldsymbol{r}},
\end{align*}
where
$$a_D=\frac{1}{(2\pi)^{\frac{D}{2}}}.$$
A counterpart for the outgoing wave is
\begin{align*}
h_0^{(1)}(kr)=2a_D\int_{\mathbb{R}^{D-1}}d\boldsymbol{k}_{\perp}\frac{e^{i\boldsymbol{k}\cdot\boldsymbol{r}}}{ k^{D-2}\sqrt{k^2-k_{\perp}^2}},
\end{align*}with $k_1=\pm \sqrt{k^2-k_{\perp}^2}$, where the sign $\pm$ is the same as the sign of $x_1$.
Now
we will use the method in \cite{31,7}.
Using the fact that the normalized hyper-spherical harmonics $C_{\boldsymbol{m}}\boldsymbol{Y}_{\boldsymbol{m}}(\boldsymbol{\theta})$ can be written as
\begin{align*}
C_{\boldsymbol{m}}\boldsymbol{Y}_{\boldsymbol{m}}(\boldsymbol{\theta})=H_{\boldsymbol{m}}\left(\frac{x_1}{r},\frac{x_2}{r},\ldots,\frac{x_D}{r}\right)
\end{align*}for some homogeneous polynomial $H_{\boldsymbol{m}}\left(x_1,\ldots,x_D\right)$ of degree $m_1=l$, we can define an operator $$\mathcal{H}_{\boldsymbol{m}}(\boldsymbol{\pa})=H_{\boldsymbol{m}}\left(\frac{\pa_{x_1}}{ik},\ldots,\frac{\pa_{x_D}}{ik}\right)$$which generalizes the operator $\mathcal{P}_{lm}$ defined in \cite{31}. It follows from definition that
\begin{align}\label{eq3_3_3}
\mathcal{H}_{\boldsymbol{m}}(\boldsymbol{\pa})e^{i\boldsymbol{k}\cdot\boldsymbol{r}}
=C_{\boldsymbol{m}}\boldsymbol{Y}_{\boldsymbol{m}}(\boldsymbol{\theta}_k)e^{i\boldsymbol{k}\cdot\boldsymbol{r}}.
\end{align}
Hence, \eqref{eq12_3_4} can be written as
\begin{equation}\label{eq3_3_2}\begin{split}
\varphi_{\boldsymbol{m}}^{\text{reg}}(\boldsymbol{x},k)=&\mathcal{C}_l^{\text{reg}}C_{\boldsymbol{m}}a_Di^l\int_{S^{D-1}_k}d\Omega_{D-1}^k\boldsymbol{Y}_{\boldsymbol{m}}(\boldsymbol{\theta}_k)
e^{i\boldsymbol{k}\cdot\boldsymbol{r}}\\
=&\mathcal{C}_l^{\text{reg}}a_Di^l\mathcal{H}_{\boldsymbol{m}}(\boldsymbol{\pa})\int_{S^{D-1}_k} d\Omega_{D-1}^ke^{i\boldsymbol{k}\cdot\boldsymbol{r}}\\
=&i^l\mathcal{C}_l^{\text{reg}}\mathcal{H}_{\boldsymbol{m}}(\boldsymbol{\pa})j_0(kr),
\end{split}\end{equation} which says that $\varphi_{\boldsymbol{m}}^{\text{reg}}(\boldsymbol{x},k)$ can be obtained by applying the operator $\mathcal{H}_{\boldsymbol{m}}(\boldsymbol{\pa})$ to $j_0(kr)$. Since $j_{\nu}(z)$ and $h_{\nu}^{(1)}(z)$ satisfies the same differential equation, it follows that
\begin{align*}
\varphi_{\boldsymbol{m}}^{\text{out}}(\boldsymbol{x},k)=&i^l\mathcal{C}_l^{\text{out}}\mathcal{H}_{\boldsymbol{m}}(\boldsymbol{\pa})h_0^{(1)}(kr).
\end{align*}
Namely,
\begin{align*}
\varphi_{\boldsymbol{m}}^{\text{out}}(\boldsymbol{x},k)=&2i^l \mathcal{C}_l^{\text{out}} a_DC_{\boldsymbol{m}}\int_{\mathbb{R}^{D-1}}d\boldsymbol{k}_{\perp}\boldsymbol{Y}_{\boldsymbol{m}}(\boldsymbol{\theta}_k)\frac{e^{i\boldsymbol{k}\cdot\boldsymbol{r}}}{ k^{D-2}\sqrt{k^2-k_{\perp}^2}}.
\end{align*}

Applying the operator $\mathcal{H}_{\boldsymbol{m}''}(\boldsymbol{\pa})$ to $\varphi_{\boldsymbol{m}}^{\text{reg}}(\boldsymbol{x},k)$ and set $\boldsymbol{x}$ equal to $\mathbf{0}$, \eqref{eq3_3_2} and \eqref{eq3_3_3} imply that
\begin{equation}\label{eq3_3_4}\begin{split}
\left[\mathcal{H}_{\boldsymbol{m}''}(\boldsymbol{\pa})\varphi_{\boldsymbol{m}}^{\text{reg}}\right](\boldsymbol{0},k)=&\mathcal{C}_l^{\text{reg}}C_{\boldsymbol{m}}
C_{\boldsymbol{m}''}a_Di^l\int_{S^{D-1}_k}d\Omega_{D-1}^k\boldsymbol{Y}_{\boldsymbol{m}}(\boldsymbol{\theta}_k)
\boldsymbol{Y}_{\boldsymbol{m}''}(\boldsymbol{\theta}_k)\\=&\mathcal{C}_l^{\text{reg}}a_Di^l\delta_{\boldsymbol{m}^*,\boldsymbol{m}''}.
\end{split}\end{equation}
For $\boldsymbol{m}=(m_1,\ldots,m_{D-1})$, we define $\boldsymbol{m}^*=(m_1,\ldots,m_{D-2},-m_{D-1})$, so that
$$\boldsymbol{Y}_{\boldsymbol{m}^*}(\boldsymbol{\theta}_k)=\boldsymbol{Y}_{\boldsymbol{m}}(\boldsymbol{\theta}_k)^*.$$

Hence, by applying $\mathcal{H}_{\boldsymbol{m}^*}(\boldsymbol{\pa})$ to \eqref{eq12_3_5} and setting $\boldsymbol{x}=\mathbf{0}$, we find that
\begin{equation}\label{eq3_4_2}
\begin{split}
G^1_{\boldsymbol{m},\boldsymbol{m}'}=&\frac{1}{C_l^{\text{reg}}a_Di^l}\left[\mathcal{H}_{\boldsymbol{m}^*}(\boldsymbol{\pa})\varphi_{\boldsymbol{m}'}^{\text{out}}\right](-\boldsymbol{L},k)\\
=&\pi(-1)^{l'}i^{D-1}C_{\boldsymbol{m}}C_{\boldsymbol{m}'}\int_{\mathbb{R}^{D-1}}d\boldsymbol{k}_{\perp}\boldsymbol{Y}_{\boldsymbol{m}^*}(\boldsymbol{\theta}_k)
\boldsymbol{Y}_{\boldsymbol{m}'}(\boldsymbol{\theta}_k)\frac{e^{-i\boldsymbol{k}\cdot\boldsymbol{L}}}{ k^{D-2}\sqrt{k^2-k_{\perp}^2}}.
\end{split}\end{equation}
In principle, one can express $\boldsymbol{Y}_{\boldsymbol{m}}(\boldsymbol{\theta}_k)
\boldsymbol{Y}_{\boldsymbol{m}'}(\boldsymbol{\theta}_k)$ as a linear combinations of $\boldsymbol{Y}_{\boldsymbol{m}''}(\boldsymbol{\theta}_k)$:
\begin{align*}
\boldsymbol{Y}_{\boldsymbol{m}}(\boldsymbol{\theta}_k)
\boldsymbol{Y}_{\boldsymbol{m}'}(\boldsymbol{\theta}_k)=\sum_{\boldsymbol{m}''}H_{\boldsymbol{m},\boldsymbol{m}'}^{\boldsymbol{m}''}\boldsymbol{Y}_{\boldsymbol{m}''}(\boldsymbol{\theta}_k).
\end{align*}When $D=3$, the constants $H_{\boldsymbol{m},\boldsymbol{m}'}^{\boldsymbol{m}''}$ are well-known and can be expressed as $3j$-symbols. However, the computations of these constants are not simple tasks. Therefore, we will use an alternative approach.

As in \cite{1}, we express the integration over $\boldsymbol{k}_{\perp}$ in polar coordinates
\begin{equation*}
\begin{split}
k_2=&k_{\perp}\cos\theta_2^k,\\
k_3=&k_{\perp}\sin\theta_2^k\cos\theta_3^k,\\
&\vdots\\
k_{D-1}=&k_{\perp}\sin\theta_2^k\ldots\sin\theta_{D-2}^k\cos\theta_{D-1}^k,\\
k_{D}=&k_{\perp}\sin\theta_2^k\ldots\sin\theta_{D-2}^k\sin\theta_{D-1}^k.
\end{split}
\end{equation*}Replacing $k$ with $i\kappa$, we have
\begin{equation*}
\begin{split}
G^1_{\boldsymbol{m},\boldsymbol{m}'}=&\pi(-1)^{l'}i^{D-1}C_{\boldsymbol{m}}C_{\boldsymbol{m}'}\int_{0}^{\infty}d\boldsymbol{k}_{\perp} \boldsymbol{Y}_{\boldsymbol{m}^*}(\boldsymbol{\theta}_k)
\boldsymbol{Y}_{\boldsymbol{m}'}(\boldsymbol{\theta}_k)\frac{e^{-i\boldsymbol{k}\cdot\boldsymbol{L}}}{ k^{D-2}\sqrt{k^2-k_{\perp}^2}}\\
=&\pi(-1)^{l} i^{-m_2-m_2'} C_{\boldsymbol{m}}^2 \int_{0}^{\infty}dk_{\perp}k_{\perp}^{D-2}\int_0^{\pi}d\theta_2^k\sin^{D-3}\theta_2^k\ldots\int_0^{\pi}d\theta_{D-2}^k\sin\theta_{D-2}^k\int_{-\pi}^{\pi}d\theta_{D-1}^k
e^{-im_{D-1}\theta_{D-1}^k+im_{D-1}'\theta_{D-1}^k}\\
&\times \frac{e^{-L\sqrt{\kappa^2+k_{\perp}^2}}}{\kappa^{D-2}\sqrt{\kappa^2+k_{\perp}^2}}\left(\frac{k_{\perp}}{\kappa}\right)^{m_2+m_2'}
C_{l-m_2}^{m_{2}+\frac{D-2}{2}}\left(\frac{\sqrt{\kappa^2+k_{\perp}^2}}{\kappa}\right)C_{l'-m_2'}^{m_{2}'+\frac{D-2}{2}}\left(\frac{\sqrt{\kappa^2+k_{\perp}^2}}{\kappa}\right)
\\&\times\prod_{j=2}^{D-2}\sin^{|m_{j+1}|}\theta_j^k C_{m_j-|m_{j+1}|}^{|m_{j+1}|+\frac{D-j-1}{2}}(\cos\theta_j^k)\prod_{j'=2}^{D-2}\sin^{|m_{j'+1}'|}\theta_{j'}^k C_{m_{j'}'-|m_{j'+1}'|}^{|m_{j'+1}'|+\frac{D-j'-1}{2}}(\cos\theta_{j'}^k).
\end{split}
\end{equation*}Integrating over $\theta_2^k,\ldots,\theta_{D-1}^k$ using the orthogonality relation \eqref{eq12_3_11} gives
\begin{equation*}\begin{split}G^1_{\boldsymbol{m},\boldsymbol{m}'}=&(-1)^{l+m_2} \delta_{\boldsymbol{m}_{\perp},\boldsymbol{m}_{\perp}'}2^{2m_2+D-3}   \Gamma\left(m_2+\frac{D-2}{2}\right)^2
\sqrt{\frac{\left(l+\frac{D-2}{2}\right)\left(l'+\frac{D-2}{2}\right)(l-m_2)!(l'-m_2)!}{(l+m_2+D-3)!(l'+m_2+D-3)!}}
\\&\times\int_{0}^{\infty}dk_{\perp}\,
\frac{e^{-L\sqrt{\kappa^2+k_{\perp}^2}}}{ \sqrt{\kappa^2+k_{\perp}^2}}\left(\frac{k_{\perp}}{\kappa}\right)^{2m_2 +D-2}
C_{l-m_2}^{m_{2}+\frac{D-2}{2}}\left(\frac{\sqrt{\kappa^2+k_{\perp}^2}}{\kappa}\right)C_{l'-m_2}^{m_{2}+\frac{D-2}{2}}\left(\frac{\sqrt{\kappa^2+k_{\perp}^2}}{\kappa}\right).
\end{split}
\end{equation*}Here $\boldsymbol{m}_{\perp}=(m_2,\ldots,m_{D-1})$. Making a change of variables $k_{\perp}=\kappa\sinh\theta$ gives
\begin{equation*}\begin{split}G^1_{\boldsymbol{m},\boldsymbol{m}'}=&(-1)^{l+m_2}\delta_{\boldsymbol{m}_{\perp},\boldsymbol{m}_{\perp}'} 2^{2m_2+D-3}   \Gamma\left(m_2+\frac{D-2}{2}\right)^2
\sqrt{\frac{\left(l+\frac{D-2}{2}\right)\left(l'+\frac{D-2}{2}\right)(l-m_2)!(l'-m_2)!}{(l+m_2+D-3)!(l'+m_2+D-3)!}}
\\&\times\int_{0}^{\infty}d\theta
 \left(\sinh\theta\right)^{2m_2 +D-2}
C_{l-m_2}^{m_{2}+\frac{D-2}{2}}\left(\cosh\theta\right)C_{l'-m_2}^{m_{2}+\frac{D-2}{2}}\left(\cosh\theta\right)e^{-\kappa L\cosh\theta}.
\end{split}
\end{equation*}
In the same way, we find that
\begin{equation*}\begin{split}G^2_{\boldsymbol{m}',\boldsymbol{m}}=&(-1)^{l+m_2}\delta_{\boldsymbol{m}_{\perp},\boldsymbol{m}_{\perp}'} 2^{2m_2+D-3}    \Gamma\left(m_2+\frac{D-2}{2}\right)^2
\sqrt{\frac{\left(l+\frac{D-2}{2}\right)\left(l'+\frac{D-2}{2}\right)(l-m_2)!(l'-m_2)!}{(l+m_2+D-3)!(l'+m_2+D-3)!}}
\\&\times\int_{0}^{\infty}d\theta
 \left(\sinh\theta\right)^{2m_2 +D-2}
C_{l-m_2}^{m_{2}+\frac{D-2}{2}}\left(\cosh\theta\right)C_{l'-m_2}^{m_{2}+\frac{D-2}{2}}\left(\cosh\theta\right)e^{-\kappa L\cosh\theta}.
\end{split}
\end{equation*}
Finally, the Casimir
 interaction energy between the two spheres are given by
\begin{align}\label{eq12_3_13}
E_{\text{Cas}}=\frac{\hbar c}{2\pi}\int_0^{\infty} d \kappa \text{Tr}\,\ln\left(1-\mathbb{M}(\kappa)\right),
\end{align}
where the  $(\boldsymbol{m},\boldsymbol{m}')$  element of $\mathbb{M}$ is given by
\begin{align*}
M_{\boldsymbol{m},\boldsymbol{m}'}=&T_{\boldsymbol{m}}^1\sum_{\boldsymbol{m}''}G^1_{\boldsymbol{m},\boldsymbol{m}''}T_{\boldsymbol{m}''}^2G_{\boldsymbol{m}'',\boldsymbol{m}'}^2.
\end{align*}

Since $\mathbb{M}$ is diagonal in $\boldsymbol{m}_{\perp}=(m_2,\ldots,m_{D-1})$, we can simplify the trace in \eqref{eq12_3_13} as follows. When $D\geq 5$,
\begin{align}\label{eq12_4_1}
 \sum_{m_3=0}^{m_2}\ldots\sum_{m_{D-2}=0}^{m_{D-3}}\sum_{m_{D-1}=-m_{D-2}}^{m_{D-2}}1= \frac{(2m_2+D-3)(m_2+D-4)!}{(D-3)!m_2!}.
\end{align}When $D=4$,
\begin{align*}
\sum_{m_3=-m_2}^{m_2}1=2m_2+1,
\end{align*}
which is equal to the right hand side of \eqref{eq12_4_1} when $D=4$.
Hence, when $D\geq 4$, the Casimir interaction energy \eqref{eq12_3_13} can be rewritten as
\begin{align}\label{eq12_3_14}
E_{\text{Cas}}=\frac{\hbar c}{2\pi}\int_0^{\infty} d \kappa \sum_{m=0}^{\infty}\frac{(2m+D-3)(m+D-4)!}{(D-3)!m!}\text{Tr}\,\ln\left(1-\mathbb{M}_{m}(\kappa)\right),
\end{align}
where $m=m_2$ and the elements $M_{m;l,l'}$ of $\mathbb{M}_{m}$ is
\begin{align*}
M_{m;l,l'}=&T_{l}^1 \sum_{l''=m}^{\infty}G^1_{m;l,l''}T_{l''}^2G_{m;l'',l'}^2.
\end{align*}
  For fixed $m$, $l,l'$ ranges from $m$ to $\infty$.

When $D=3$, we can also represent the Casimir interaction energy by \eqref{eq12_3_14} provided that the summation $\sum_{m=0}^{\infty}$ is replaced by the summation $\sum_{m=0}^{\infty}\!'$, where the prime $\prime$ indicates that the term $m=0$ is summed with weight $1/2$.
\section{Large separation asymptotic behavior}

In this section, we consider the large separation asymptotic behavior of the Casimir interaction energy.
Expanding the logarithm in \eqref{eq12_3_14}, we have
\begin{align}\label{eq12_3_14_2}
E_{\text{Cas}}=-\frac{\hbar c}{2\pi}\sum_{s=0}^{\infty}\frac{1}{s+1}\int_0^{\infty} d \kappa \sum_{m=0}^{\infty}\frac{(2m+D-3)(m+D-4)!}{(D-3)!m!}\sum_{l_0=m}^{\infty}
\sum_{l_1=m}^{\infty}\ldots\sum_{l_s=m}^{\infty}\prod_{j=0}^sM_{m; l_jl_{j+1}}.
\end{align}
Making the change of variables $$\kappa\mapsto\frac{\kappa}{L},$$ we find from the definition of $M_{m;l,l'}$ that in order to obtain the leading asymptotic behavior of the Casimir interaction energy when $L\gg 1$, we need the following asymptotic behaviors when $z\rightarrow 0$:
\begin{align*}
I_{\nu}(z)\sim & \frac{1}{\Gamma(\nu+1)}\left(\frac{z}{2}\right)^{\nu},\\
K_{\nu}(z)\sim &\frac{\Gamma(\nu)}{2}\left(\frac{z}{2}\right)^{-\nu}.
\end{align*}
 It follows that for $i=1, 2$,
\begin{equation}\label{eq12_4_7}\begin{split}
T_{l}^{i, \text{D}}\left(\frac{\kappa}{L}\right)\sim & \frac{1}{2^{2l+D-3}\Gamma\left(l+\frac{D-2}{2}\right)\Gamma\left(l+\frac{D}{2}\right)}\left(\frac{\kappa R_i}{L}\right)^{2l+D-2},\\
T_{l}^{i, \text{N}}\left(\frac{\kappa}{L}\right)\sim & -\frac{l}{l+D-2}\frac{1}{2^{2l+D-3}\Gamma\left(l+\frac{D-2}{2}\right)\Gamma\left(l+\frac{D}{2}\right)}\left(\frac{\kappa R_i}{L}\right)^{2l+D-2},\quad\text{if}\;l\neq 0,\\
T_{0}^{i,\text{N}}\left(\frac{\kappa}{L}\right)\sim &-\frac{1}{2^{D-1}\Gamma\left(\frac{D}{2}\right)\Gamma\left(\frac{D+2}{2}\right)}\left(\frac{\kappa R_i}{L}\right)^{D}.
\end{split}\end{equation}
 From these asymptotic behaviors, we find that the leading contribution to the large separation asymptotic behavior of the Casimir interaction energy comes from lower   $l$ (and hence lower $m$) as well as smallest possible $s$, i.e., $s=0$. For Dirichlet boundary conditions, we only take $l=0$ (and hence $m=0$) for the leading term of large separation asymptotic expansion. However, for Neumann boundary conditions, both the $l=0$ and $l=1$ terms of $T_{l}^{i, \text{N}}\left(\kappa/L\right)$ have the same order in $L$, so we have to take both $l=0$ and $l=1$ terms to compute  the  leading term of the large separation asymptotic expansion. We have
\begin{equation}
\begin{split}
E_{\text{Cas}}^{\text{DD}}\sim &-\frac{\hbar c}{2\pi L}\int_0^{\infty}d\kappa T_0^{1,\text{D}}G_{0;00}^1T_{0}^{2,\text{D}}G_{0;00}^2,
\\
E_{\text{Cas}}^{\text{DN}}\sim &-\frac{\hbar c}{2\pi L}\int_0^{\infty}d\kappa \left(T_0^{1,\text{D}}G_{0;00}^1T_{0}^{2,\text{N}}G_{0;00}^2+T_0^{1,\text{D}}G_{0;01}^1T_{1}^{2,\text{N}}G_{0;10}^2\right),\\
E_{\text{Cas}}^{\text{NN}}\sim & -\frac{\hbar c}{2\pi L}\int_0^{\infty}d\kappa \left(T_0^{1,\text{N}}G_{0;00}^1T_{0}^{2,\text{N}}G_{0;00}^2+T_0^{1,\text{N}}G_{0;01}^1T_{1}^{2,\text{N}}G_{0;10}^2+T_1^{1,\text{N}}G_{0;10}^1T_{0}^{2,\text{N}}G_{0;01}^2
\right.\\&\hspace{4cm}\left.+T_1^{1,\text{N}}G_{0;11}^1T_{1}^{2,\text{N}}G_{0;11}^2+(D-1)T_1^{1,\text{N}}G_{1;11}^1T_{1}^{2,\text{N}}G_{1;11}^2\right).
\end{split}
\end{equation}
Using
\begin{align*}
C_0^{\nu}(z)=1,\hspace{1cm}C_1^{\nu}(z)=2\nu z,
\end{align*}we find that
\begin{equation*}\begin{split}
G_{0;00}^1=G_{0;00}^2=&\sqrt{\pi}\frac{\Gamma\left(\frac{D}{2}\right)}{\Gamma\left(\frac{D-1}{2}\right)}\int_0^{\infty}d\theta\,\sinh^{D-2}\theta e^{-\kappa\cosh\theta}\\=& \left(\frac{2}{\kappa}\right)^{\frac{D-2}{2}}
\Gamma\left(\frac{D}{2}\right)K_{\frac{D-2}{2}}\left(\kappa\right),\end{split}\end{equation*}
\begin{equation*}\begin{split}
G_{0;01}^1=G_{0;10}^2=-G_{0;10}^1=-G_{0;01}^2=&\sqrt{\pi D}\frac{\Gamma\left(\frac{D}{2}\right)}{\Gamma\left(\frac{D-1}{2}\right)}\int_0^{\infty}d\theta\,\sinh^{D-2}\theta\cosh\theta e^{-\kappa\cosh\theta}\\=&\sqrt{D} \left(\frac{2}{\kappa}\right)^{\frac{D-2}{2}}\Gamma\left(\frac{D}{2}\right)K_{\frac{D}{2}}\left(\kappa\right),
\end{split}\end{equation*}
\begin{equation*}\begin{split}
G_{0;11}^1=G_{0;11}^2=& -2\sqrt{\pi  }\frac{\Gamma\left(\frac{D+2}{2}\right)}{\Gamma\left(\frac{D-1}{2}\right)}\int_0^{\infty}d\theta\,\sinh^{D-2}\theta\cosh^2\theta e^{-\kappa\cosh\theta}\\
=&-2\Gamma\left(\frac{D+2}{2}\right)\left(\left(\frac{2}{\kappa}\right)^{\frac{D-2}{2}}
K_{\frac{D-2}{2}}\left(\kappa\right)+\frac{D-1}{2}\left(\frac{2}{\kappa}\right)^{\frac{D}{2}}
K_{\frac{D}{2}}\left(\kappa\right)\right),
\end{split}\end{equation*}
\begin{equation*}\begin{split}
G_{1;11}^1=G_{1;11}^2=&\sqrt{\pi}\frac{\Gamma\left(\frac{D+2}{2}\right)}{\Gamma\left(\frac{D+1}{2}\right)}\int_0^{\infty}d\theta\,\sinh^{D}\theta e^{-\kappa\cosh\theta}\\=& \left(\frac{2}{\kappa}\right)^{\frac{D}{2}}
\Gamma\left(\frac{D+2}{2}\right)K_{\frac{D}{2}}\left(\kappa\right).\end{split}\end{equation*}
It follows that the leading term of the large separation asymptotic expansion of the Casimir interaction energy is given by
\begin{equation*}
\begin{split}
E_{\text{Cas}}^{\text{DD}}\sim & -\frac{\hbar c}{2\sqrt{\pi}}\frac{\Gamma\left(\frac{D-1}{2}\right)^2\Gamma\left(D-\frac{3}{2}\right)}{\Gamma\left(\frac{D-2}{2}\right)^2(D-2)!}
\frac{(R_1R_2)^{D-2}}{L^{2D-3}},\\
E_{\text{Cas}}^{\text{DN}}\sim & \frac{\hbar c}{4\sqrt{\pi}}\frac{\Gamma\left(\frac{D+1}{2}\right)^2\Gamma\left(D-\frac{1}{2}\right)}{\Gamma\left(\frac{D-2}{2}\right)\Gamma\left(\frac{D+2}{2}\right)D!}\frac{2D^2-1}{D-1}
\frac{R_1^{D-2}R_2^{D}}{L^{2D-1}},\\
E_{\text{Cas}}^{\text{ND}}\sim & \frac{\hbar c}{4\sqrt{\pi}}\frac{\Gamma\left(\frac{D+1}{2}\right)^2\Gamma\left(D-\frac{1}{2}\right)}{\Gamma\left(\frac{D-2}{2}\right)\Gamma\left(\frac{D+2}{2}\right)D!}\frac{2D^2-1}{D-1}
\frac{R_1^{D}R_2^{D-2}}{L^{2D-1}},\\
E_{\text{Cas}}^{\text{NN}}\sim & -\frac{\hbar c}{8\sqrt{\pi}}\frac{\Gamma\left(\frac{D+3}{2}\right)^2\Gamma\left(D+\frac{1}{2}\right)}{\Gamma\left(\frac{D+2}{2}\right)^2 (D+2)!}\frac{(2D^2+2D-1)(2D^2+2D-3)}{(D-1)^2}
\frac{(R_1R_2)^{D}}{L^{2D+1}}.
\end{split}
\end{equation*}
In other words, the large separation leading terms of the Casimir interaction energies are of order $L^{-2D+3}$, $L^{-2D+1}$ and $L^{-2D-1}$ respectively for DD, DN/ND and NN boundary conditions. Hence, when the separation between the spheres is large, the interaction is strongest for Dirichlet-Dirchlet boundary conditions, and weakest for Neumann-Neumann boundary conditions. Moreover, the interaction gets weaker for higher dimensions.

When $D=3$, we find that
\begin{align*}
E_{\text{Cas}}^{\text{DD}}\sim &-\frac{\hbar cR_1R_2}{4\pi L^3},\\
E_{\text{Cas}}^{\text{DN}}\sim &\frac{17\hbar c R_1R_2^3}{48\pi L^5},\\
E_{\text{Cas}}^{\text{NN}}\sim &\frac{17 \hbar cR_1^3R_2}{48\pi L^5},\\
E_{\text{Cas}}^{\text{NN}}\sim &-\frac{161\hbar c R_1^3R_2^3}{96\pi L^7},
\end{align*}which agree with the results derived in \cite{6}.

\section{Proximity force approximation and small separation asymptotic behavior}
The proximity force approximation approximates the Casimir interaction energy between two objects by summing  the local Casimir energy density between two planes over the surfaces. For two planes both with Dirichlet (D) or Neumann (N) boundary conditions in $(D+1)$-dimensional Minkowski spacetime, the Casimir energy density is
\begin{align*}
\mathcal{E}_{\text{Cas}}^{\parallel,\text{DD/NN}} (d)=-\hbar c\frac{\Gamma\left(\frac{D+1}{2}\right)\zeta(D+1)}{2^{D+1}\pi^{\frac{D+1}{2}}}\frac{1}{d^{D}}=\frac{b_D^{\text{DD/NN}}}{d^{D}};
\end{align*}whereas if one plane is Dirichlet and one is Neumann, the Casimir energy density is
\begin{align*}
\mathcal{E}_{\text{Cas}}^{\parallel,\text{DN/ND}} (d)=\hbar c\left(1-2^{-D}\right)\frac{\Gamma\left(\frac{D+1}{2}\right)\zeta(D+1)}{2^{D+1}\pi^{\frac{D+1}{2}}}\frac{1}{d^{D}}=\frac{b_D^{\text{DN/ND}}}{d^{D}}.
\end{align*}Here $\zeta(z)=\sum_{n=1}^{\infty}n^{-z}$ is the Riemann zeta function.

Let $(R_1,\theta_1,\ldots,\theta_{D-1})$ be a point on the sphere with radius $R_1$ in hyper-spherical coordinates, the distance from this point to the sphere with radius $R_2$ is
\begin{equation*}
d(\boldsymbol{\theta})=\sqrt{L^2-2R_1L\cos\theta_1+R_1^2}-R_2.
\end{equation*}Notice that this only depends on $\theta_1$.

The proximity force approximation to the Casimir interaction energy between the two spheres is given by
\begin{equation*}
\begin{split}
E_{\text{Cas}}^{\text{PFA}}=&R_1^{D-1}\int_0^{\pi}d\theta_1\sin^{D-2}\theta_1\int_0^{\pi}d\theta_2\sin^{D-3}\theta_2\ldots\int_0^{\pi}d\theta_{D-2}\sin\theta_{D-2}\int_{-\pi}^{\pi}
d\theta_{D-1} \mathcal{E}_{\text{Cas}}^{\parallel}\left(d\left(\boldsymbol{\theta}\right)\right)\\
=&\frac{2\pi^{\frac{D-1}{2}}}{\Gamma\left(\frac{D-1}{2}\right)}b_DR_1^{D-1}\int_0^{\pi}d\theta_1\frac{\sin^{D-2}\theta_1}{\left(\sqrt{L^2-2R_1L\cos\theta_1+R_1^2}-R_2\right)^D}.
\end{split}
\end{equation*}
Let
\begin{equation*}
u=\sqrt{L^2-2R_1L\cos\theta_1+R_1^2}-R_2.
\end{equation*}
Then
\begin{equation*}\begin{split}
E_{\text{Cas}}^{\text{PFA}}=&\frac{2\pi^{\frac{D-1}{2}}}{\Gamma\left(\frac{D-1}{2}\right)}b_D\frac{R_1 }{2^{D-3} L^{D-2}}\int_d^{L+R_1-R_2}du\left(u+R_2\right)
u^{-D}\\&\hspace{4cm}\times\left(\left[u-d\right]\left[u+R_2+L-R_1\right]\left[u+L+R_1+R_2\right]\left[L+R_1-R_2-u\right]\right)^{\frac{D-3}{2}}\\
\sim & \frac{(2\pi)^{\frac{D-1}{2}}}{\Gamma\left(\frac{D-1}{2}\right)} \left(\frac{R_1R_2}{R_1+R_2}\right)^{\frac{D-1}{2}}\frac{b^D}{d^{\frac{D+1}{2}}}\int_1^{\infty} dv v^{-D}(v-1)^{\frac{D-3}{2}}\\
=&\frac{(2\pi)^{\frac{D-1}{2}}\Gamma\left(\frac{D+1}{2}\right)}{\Gamma(D)} \left(\frac{R_1R_2}{R_1+R_2}\right)^{\frac{D-1}{2}}\frac{b^D}{d^{\frac{D+1}{2}}}\\
=&\frac{\pi^{\frac{D}{2}}}{2^{\frac{D-1}{2}}\Gamma\left(\frac{D}{2}\right)} \left(\frac{R_1R_2}{R_1+R_2}\right)^{\frac{D-1}{2}}\frac{b^D}{d^{\frac{D+1}{2}}}.
\end{split}\end{equation*}
Here $d=L-R_1-R_2$ is the distance between the two spheres. Hence, we find that the proximity force approximation to the Casimir interaction energy between two spheres is
\begin{equation}\label{eq12_5_10}\begin{split}
E_{\text{Cas}}^{\text{PFA, DD/NN}}\sim &-\frac{\Gamma\left(\frac{D+1}{2}\right)\zeta(D+1)}{2^{\frac{3D+1}{2}}\sqrt{\pi}\Gamma\left(\frac{D}{2}\right)}\left(\frac{R_1R_2}{R_1+R_2}\right)^{\frac{D-1}{2}} \frac{\hbar c}{d^{\frac{D+1}{2}}},\\
E_{\text{Cas}}^{\text{PFA, DN/ND}}\sim &(1-2^{-D})\frac{\Gamma\left(\frac{D+1}{2}\right)\zeta(D+1)}{2^{\frac{3D+1}{2}}\sqrt{\pi}\Gamma\left(\frac{D}{2}\right)}\left(\frac{R_1R_2}{R_1+R_2}\right)^{\frac{D-1}{2}} \frac{\hbar c}{d^{\frac{D+1}{2}}}.
\end{split}\end{equation}

Next, we derive the small separation asymptotic behavior of the Casimir interaction energy from the TGTG formula. Set
$$\tilde{m}=m+\frac{D-3}{2},$$and make a change of variables
$$\kappa =\frac{\omega}{R_1+R_2},$$we obtain from \eqref{eq12_3_14} that
 \begin{align}\label{eq3_6_1}
E_{\text{Cas}}=-\frac{\hbar c}{\pi(R_1+R_2)}\frac{1}{(D-3)!}\sum_{s=0}^{\infty}\frac{1}{s+1}\int_0^{\infty} d \omega \sum_{m=0}^{\infty}\frac{\tilde{m}\left(\tilde{m}+\frac{D-5}{2}\right)!}{ \left(\tilde{m}-\frac{D-3}{2}\right)!}  \sum_{l_0=\tilde{m}-\frac{D-3}{2}}^{\infty}\sum_{l_1=\tilde{m}-\frac{D-3}{2}}^{\infty}\ldots \sum_{l_s=\tilde{m}-\frac{D-3}{2}}^{\infty}
\prod_{j=0}^s M_{l_j,l_{j+1}},
\end{align}
where
\begin{align}\label{eq3_10_1}
M_{l_j,l_{j+1}}=T_{l_j}^1\sum_{l_j'=\tilde{m}}^{\infty}U^1_{l_j,l_j'}T_{l_j'}^2U^2_{l_j',l_{j+1}},
\end{align}
\begin{equation*}
\begin{split}
&T_{l_j}^{1, \text{D}}  =\frac{I_{l_j+\frac{D-2}{2}}(\omega a_1)}{K_{l_j+\frac{D-2}{2}}(\omega a_1)},\\
& T_{l_j}^{1, \text{N}} =\frac{-\frac{D-2}{2}I_{l_j+\frac{D-2}{2}}(\omega a_1)+\omega a_1I_{l_j+\frac{D-2}{2}}'(\omega a_1)}
{-\frac{D-2}{2}K_{l_j+\frac{D-2}{2}}(\omega a_1)+\omega a_1 K_{l_j+\frac{D-2}{2}}'(\omega a_1)},\\
&T_{l_j'}^{2, \text{D}}  =\frac{I_{l_j'+\frac{D-2}{2}}(\omega a_2)}{K_{l_j'+\frac{D-2}{2}}(\omega a_2)},\\
& T_{l_j'}^{2, \text{N}} =\frac{-\frac{D-2}{2}I_{l_j'+\frac{D-2}{2}}(\omega a_2)+\omega a_2I_{l_j'+\frac{D-2}{2}}'(\omega a_2)}
{-\frac{D-2}{2}K_{l_j'+\frac{D-2}{2}}(\omega a_2)+\omega a_2 K_{l_j'+\frac{D-2}{2}}'(\omega a_2)},
\end{split}
\end{equation*}
\begin{equation*}\begin{split}U^1_{l_j,l_j'}=& 2^{2\tilde{m}}   \Gamma\left(\tilde{m}+\frac{1}{2}\right)^2
\sqrt{\frac{\left(l_j+\frac{D-2}{2}\right)\left(l_j'+\frac{D-2}{2}\right)\Gamma\left(l_j+\frac{D-1}{2}-\tilde{m}\right)\Gamma\left(l_j'+\frac{D-1}{2}-\tilde{m}\right)}
{\Gamma\left(l_j+\frac{D-1}{2}+\tilde{m}\right)\Gamma\left(l_j'+\frac{D-1}{2}+\tilde{m}\right)}}
\\&\times\int_{0}^{\infty}d\theta\sinh\theta
 \left(\sinh\theta\right)^{2\tilde{m}}
C_{l_j+\frac{D-3}{2}-\tilde{m}}^{\tilde{m}+\frac{1}{2}}\left(\cosh\theta\right)C_{l_j'+\frac{D-3}{2}-\tilde{m}}^{\tilde{m}+\frac{1}{2}}
\left(\cosh\theta\right)e^{-\omega\left(1+\vep\right)\cosh\theta},\\
U^2_{l_{j+1},l_j'}=& 2^{2\tilde{m}}   \Gamma\left(\tilde{m}+\frac{1}{2}\right)^2
\sqrt{\frac{\left(l_{j+1}+\frac{D-2}{2}\right)\left(l_j'+\frac{D-2}{2}\right)\Gamma\left(l_{j+1}+\frac{D-1}{2}-\tilde{m}\right)\Gamma\left(l_j'+\frac{D-1}{2}-\tilde{m}\right)}
{\Gamma\left(l_{j+1}+\frac{D-1}{2}+\tilde{m}\right)\Gamma\left(l_j'+\frac{D-1}{2}+\tilde{m}\right)}}
\\&\times\int_{0}^{\infty}d\theta\sinh\theta
 \left(\sinh\theta\right)^{2\tilde{m}}
C_{l_{j+1}+\frac{D-3}{2}-\tilde{m}}^{\tilde{m}+\frac{1}{2}}\left(\cosh\theta\right)C_{l_j'+\frac{D-3}{2}-\tilde{m}}^{\tilde{m}+\frac{1}{2}}
\left(\cosh\theta\right)e^{-\omega\left(1+\vep\right)\cosh\theta}.
\end{split}
\end{equation*}Here
\begin{align*}
a_i=\frac{R_i}{R_1+R_2},\hspace{1cm}\vep=\frac{d}{R_1+R_2}.
\end{align*}Notice that $U^2_{l_{j+1},l_j'}$ is obtained from $U^1_{l_j,l_j'}$ by replacing $l_j$ with $l_{j+1}$.

Next, we introduce new variables $n_1, \ldots, n_s$, $q_0, q_1, \ldots, q_s$ and $\tau$ such that
\begin{align*}
l_0=&l,\hspace{1cm} l_j=l+n_j,\quad 1\leq j\leq s, \\
l_j'=&\frac{a_2}{2a_1}\left(l_j+l_{j+1}\right)+q_j=\frac{a_2}{a_1}l+\frac{a_2}{2a_1}\left(n_j+n_{j+1}\right)+q_j,\quad 0\leq j\leq s,\\
\omega=&\frac{l\sqrt{1-\tau^2}}{a_1\tau}.
\end{align*}When $\vep$ is small, the leading contributions to the Casimir interaction energy come from terms with $l\sim \vep^{-1}$, $n_i, q_i, \tilde{m} \sim \vep^{-1/2}$ and $\tau\sim 1$.

As explained in \cite{1}, we have
\begin{equation}\label{eq3_12_1}\begin{split}
&\sinh^{\tilde{m}}\theta C_{l_j+\frac{D-3}{2}-\tilde{m}}^{\tilde{m}+\frac{1}{2}}(\cosh\theta)\\=&\frac{\Gamma\left(l_j+\frac{D-1}{2}+\tilde{m}\right)}{\Gamma\left(\tilde{m}+\frac{1}{2}\right)
\Gamma\left(l_j+\frac{D-1}{2}\right)}
 \frac{1}{ 2^{\tilde{m}}\sqrt{\pi}  }\int_{-\frac{\pi}{2}}^{\frac{\pi}{2}} d\varphi\left(\cosh\theta+\sinh\theta\cos2\varphi\right)^{l_j+\frac{D-3}{2}}e^{2i\tilde{m}\varphi}\\
 =&\frac{\Gamma\left(l_j+\frac{D-1}{2}+\tilde{m}\right)}{\Gamma\left(\tilde{m}+\frac{1}{2}\right)}\frac{1}{ 2^{\tilde{m}}\sqrt{\pi}  }
 \sum_{k=0}^{\infty} \frac{1}{k!\Gamma\left(l_j+\frac{D-1}{2}-k\right)}e^{\left(l_j+\frac{D-3}{2}-2k\right)\theta}\int_{-\frac{\pi}{2}}^{\frac{\pi}{2}}d\varphi
 \left(\cos\varphi\right)^{2l_j+D-3-2k}\left(\sin\varphi\right)^{2k}e^{2i\tilde{m}\varphi}.
\end{split}\end{equation}
Hence,
\begin{align*}
U^1_{l_j,l_j'}=&   \frac{1}{\pi}\sum_{k=0}^{\infty}\sum_{k'=0}^{\infty}\frac{1}{k!k'!}\mathcal{N}_{l_j,l_j';k,k'}\int_{0}^{\infty}d\theta\sinh\theta e^{\left(l_j+l_j'+D-3-2k-2k'\right)\theta}
 e^{-\omega\left(1+\vep\right)\cosh\theta}\\
 &\times \int_{-\frac{\pi}{2}}^{\frac{\pi}{2}}d\varphi
 \left(\cos\varphi\right)^{2l_j+D-3-2k}\left(\sin\varphi\right)^{2k}e^{2i\tilde{m}\varphi}\int_{-\frac{\pi}{2}}^{\frac{\pi}{2}}d\varphi'
 \left(\cos\varphi'\right)^{2l_j'+D-3-2k'}\left(\sin\varphi'\right)^{2k'}e^{2i\tilde{m}\varphi'},
\end{align*}where
\begin{align*}
\mathcal{N}_{l_j,l_j';k,k'}=&\sqrt{ \left(l_j+\frac{D-2}{2}\right)\left(l_j'+\frac{D-2}{2}\right)}\\&\times \frac{\sqrt{\Gamma\left(l_j+\frac{D-1}{2}-\tilde{m}\right)\Gamma\left(l_j'+\frac{D-1}{2}-\tilde{m}\right) \Gamma\left(l_j+\frac{D-1}{2}+\tilde{m}\right)\Gamma\left(l_j'+\frac{D-1}{2}+\tilde{m}\right)}}{\Gamma\left(l_j+\frac{D-1}{2}-k\right)\Gamma\left(l_j'+\frac{D-1}{2}-k'\right)}.
\end{align*}
As in \cite{1, 2}, we can now find the asymptotic behaviors of $U^1_{l_j,l_j'}$.
Using
\begin{align*}
\left(\cos\varphi\right)^{2l_j+D-3-2k}=&\exp\left(\left(2l_j+D-3-2k\right)\ln\cos\varphi\right)\\= &\exp\left(-\left(l_j+\frac{D-3}{2}-k\right)\varphi^2-
\frac{\left(2l_j+D-3-2k\right)}{12}\varphi^4+\ldots\right),\\
\left(\sin\varphi\right)^{2k}=&\varphi^{2k}\left(1-\frac{\varphi^2}{6}+\ldots\right)^{2k},
\end{align*}we observe that when $\vep$ is small,  the main contribution to the Casimir interaction energy comes from terms with $\varphi\sim \vep^{1/2}$. Making a change of variable
$$\varphi\mapsto \frac{\varphi}{\sqrt{l}}$$ so that $\varphi\sim 1$, we have an asymptotic expansion of the form:
\begin{equation*}\begin{split}\int_{-\frac{\pi}{2}}^{\frac{\pi}{2}}d\varphi
 \left(\cos\varphi\right)^{2l_j+D-3-2k}\left(\sin\varphi\right)^{2k}e^{2i\tilde{m}\varphi}\sim \frac{1}{l^{k+\frac{1}{2}}}\int_{-\infty}^{\infty}
 d\varphi \varphi^{2k}\left(1+\mathcal{B}_{j,2}\right)\exp\left(\mathcal{A}_{j,1}+\mathcal{A}_{j,2}\right)\exp\left(-\varphi^2+\frac{2i\tilde{m}}{\sqrt{l}}\varphi\right).
\end{split}\end{equation*}Here and in the following, for any $\mathcal{X}$, $\mathcal{X}_{j,1}$ and $\mathcal{X}_{j,2}$ represent respectively terms of order $\sqrt{\vep}$ and $\vep$.

In a similar way, we have an expansion of the form
\begin{equation*}\begin{split}\int_{-\frac{\pi}{2}}^{\frac{\pi}{2}}d\varphi'
 \left(\cos\varphi'\right)^{2l_j'+D-3-2k'}\left(\sin\varphi'\right)^{2k'}e^{2i\tilde{m}\varphi'}\sim \frac{1}{l^{k'+\frac{1}{2}}}\int_{-\infty}^{\infty}
 d\varphi' \varphi^{\prime 2k'}\left(1+\mathcal{D}_{j,2}\right)\exp\left(\mathcal{C}_{j,1}+\mathcal{C}_{j,2}\right)\exp\left(-\frac{a_2}{a_1}\varphi^{\prime 2}+\frac{2i\tilde{m}}{\sqrt{l}}\varphi'\right).
\end{split}\end{equation*}

Let $\theta_0>0$ be such that
\begin{align*}
\sinh\theta_0=\frac{\tau}{\sqrt{1-\tau^2}}.
\end{align*}Namely,
\begin{align*}
\theta_0=\frac{1}{2}\ln\frac{1+\tau}{1-\tau}.
\end{align*}
Then
\begin{equation*}
\begin{split}
&\int_{0}^{\infty}d\theta\sinh\theta e^{\left(l_j+l_j'+D-3-2k-2k'\right)\theta}
 e^{-\omega\left(1+\vep\right)\cosh\theta}\\=&\int_{-\theta_0}^{\infty}d\theta\sinh\left(\theta+\theta_0\right) e^{\left(l_j+l_j'+D-3-2k-2k'\right)(\theta+\theta_0)}
 e^{-\omega\left(1+\vep\right)\cosh\left(\theta+\theta_0\right)}.
\end{split}
\end{equation*}When $\vep\ll 1$, the main contribution to the Casimir interaction energy comes from terms with $\theta\sim \vep^{1/2}$. Hence up to leading contributions, the integration from $-\theta_0$ to $\infty$ can be replaced by integration from $-\infty$ to $\infty$. Since
\begin{equation*}
\begin{split}
\sinh\left(\theta+\theta_0\right) =&\frac{\tau}{\sqrt{1-\tau^2}}\left(1+\frac{\theta}{\tau}+\frac{\theta^2}{2}+\ldots\right),\\
\cosh\left(\theta+\theta_0\right) =&\frac{1}{\sqrt{1-\tau^2}}\left(1+ \tau\theta +\frac{\theta^2}{2}+\frac{\tau}{6}\theta^3+\frac{\theta^4}{24}+\ldots\right),
\end{split}
\end{equation*}
we obtain an expansion of the form:
\begin{equation*}
\begin{split}
 &\int_{-\theta_0}^{\infty}d\theta\sinh\left(\theta+\theta_0\right) e^{\left(l_j+l_j'+D-3-2k-2k'\right)(\theta+\theta_0)}
 e^{-\omega\left(1+\vep\right)\cosh\left(\theta+\theta_0\right)}\\
 \sim &\left(\frac{1+\tau}{1-\tau}\right)^{\frac{l}{2a_1}+\frac{(a_1+1)}{4a_1}n_j+\frac{a_2}{4a_1}n_{j+1}+\frac{q_j}{2}+\frac{D-3}{2}-k-k'}\frac{\tau}{\sqrt{1-\tau^2}}
 \int_{-\infty}^{\infty} d\theta \left(1+\mathcal{F}_{j,1}+\mathcal{F}_{j,2}\right)\\
 &\times\exp\left(\mathcal{E}_{j,1}+\mathcal{E}_{j,2}\right)
 \exp\left(-\frac{l}{a_1\tau}-\frac{l\theta^2}{2a_1\tau}+\left(\frac{a_1+1}{2a_1}n_j+\frac{a_2}{2a_1}n_{j+1}+q_j\right)\theta-\frac{l_1\vep}{a_1\tau}\right).
\end{split}
\end{equation*}For the term $\mathcal{N}_{l_j,l_j';k,k'}$,
\begin{align*}
\sqrt{ \left(l_j+\frac{D-2}{2}\right)\left(l_j'+\frac{D-2}{2}\right)}\sim \sqrt{\frac{a_2}{a_1}}l\left(1+\mathcal{G}_{j,1}+\mathcal{G}_{j,2}\right),
\end{align*}and by using the Stirling's formula
\begin{equation*}
\ln\Gamma(z)\sim \left(z-\frac{1}{2}\right)\ln z-z +\frac{1}{2}\ln (2\pi)+\frac{1}{12z}+\ldots,
\end{equation*}we have
\begin{equation*}
\begin{split}
&\frac{\sqrt{\Gamma\left(l_j+\frac{D-1}{2}-\tilde{m}\right)\Gamma\left(l_j'+\frac{D-1}{2}-\tilde{m}\right) \Gamma\left(l_j+\frac{D-1}{2}+\tilde{m}\right)\Gamma\left(l_j'+\frac{D-1}{2}+\tilde{m}\right)}}{\Gamma\left(l_j+\frac{D-1}{2}-k\right)\Gamma\left(l_j'+\frac{D-1}{2}-k'\right)}\\
\sim & l^{k+k'}\left(\frac{a_2}{a_1}\right)^{k_2}\exp\left(\mathcal{H}_{j,1}+\mathcal{H}_{j,2}\right)\exp\left(\frac{\tilde{m}^2}{2a_2l}\right)
\end{split}
\end{equation*}
Hence,
\begin{equation*}
\begin{split}
U^1_{l_j,l_j'}\sim &   \frac{1}{\pi}\sum_{k=0}^{\infty}\sum_{k'=0}^{\infty}\frac{1}{k!k'!}\left(\frac{a_2}{a_1}\right)^{k'+\frac{1}{2}}\left(\frac{1+\tau}{1-\tau}\right)^{\frac{l}{2a_1}+\frac{(a_1+1)}{4a_1}n_j+\frac{a_2}{4a_1}n_{j+1}+\frac{q_j}{2}+\frac{D-3}{2}-k-k'}\frac{\tau}{\sqrt{1-\tau^2}}
 \exp\left(\frac{\tilde{m}^2}{2a_2l}\right)\\&\times \int_{-\infty}^{\infty} d\theta
 \exp\left(-\frac{l}{a_1\tau}-\frac{l\theta^2}{2a_1\tau}+\left(\frac{a_1+1}{2a_1}n_j+\frac{a_2}{2a_1}n_{j+1}+q_j\right)\theta
 -\frac{l_1\vep}{a_1\tau}\right)\\&\times \int_{-\infty}^{\infty}d\varphi \varphi^{2k}\exp\left(-\varphi^2+\frac{2i\tilde{m}}{\sqrt{l}}\varphi\right)
 \int_{-\infty}^{\infty}d\varphi' \varphi^{\prime 2k'}\exp\left(-\frac{a_2}{a_1}\varphi^{\prime 2}+\frac{2i\tilde{m}}{\sqrt{l}}\varphi'\right)\left(1+\mathcal{J}_{j,1}+\mathcal{J}_{j,2}\right),
\end{split}\end{equation*}
where
\begin{align*}
\mathcal{I}_{j,1}=&\mathcal{A}_{j,1}+\mathcal{C}_{j,1}+\mathcal{E}_{j,1}+\mathcal{H}_{j,1},\\
\mathcal{I}_{j,2}=&\mathcal{A}_{j,2}+\mathcal{C}_{j,2}+\mathcal{E}_{j,2}+\mathcal{H}_{j,2},\\
\mathcal{J}_{j,1}=&\mathcal{F}_{j,1}+\mathcal{G}_{j,1}+\mathcal{I}_{j,1},\\
\mathcal{J}_{j,2}=&\mathcal{F}_{j,1}\mathcal{G}_{j,1}+\mathcal{F}_{j,1}\mathcal{I}_{j,1}+\mathcal{G}_{j,1}\mathcal{I}_{j,1}+\mathcal{B}_{j,2}+\mathcal{D}_{j,2}+\mathcal{F}_{j,2}+\mathcal{G}_{j,2}
+\mathcal{I}_{j,2}+\frac{1}{2}\mathcal{I}_{j,1}^2.
\end{align*}
Now we perform the summation over $k$ and $k'$ using the following identities:
\begin{align*}
&\sum_{k=0}^{\infty} \frac{v^k}{k!}=e^v,\\
&\sum_{k=0}^{\infty} k\frac{v^k}{k!}=ve^v,\\
&\sum_{k=0}^{\infty} k^2\frac{v^k}{k!}=(v^2+v)e^v.
\end{align*}
This gives an expansion of the form
\begin{equation}\label{eq3_10_2}
\begin{split}
U^1_{l_j,l_j'}\sim &   \frac{1}{\pi}\sqrt{\frac{a_2}{a_1}}
\left(\frac{1+\tau}{1-\tau}\right)^{\frac{l}{2a_1}+\frac{(a_1+1)}{4a_1}n_j+\frac{a_2}{4a_1}n_{j+1}+\frac{q_j}{2}+\frac{D-3}{2} }\frac{\tau}{\sqrt{1-\tau^2}}\exp\left(\frac{\tilde{m}^2}{2a_2l}\right)
 \\&\times \int_{-\infty}^{\infty} d\theta
 \exp\left(-\frac{l}{a_1\tau}-\frac{l\theta^2}{2a_1\tau}+\left(\frac{a_1+1}{2a_1}n_j+\frac{a_2}{2a_1}n_{j+1}+q_j\right)\theta
 -\frac{l_1\vep}{a_1\tau}\right)\\&\times \int_{-\infty}^{\infty}d\varphi  \exp\left(-\frac{2\tau}{1+\tau}\varphi^2+\frac{2i\tilde{m}}{\sqrt{l}}\varphi\right)
 \int_{-\infty}^{\infty}d\varphi'  \exp\left(-\frac{a_2}{a_1}\frac{2\tau}{1+\tau}\varphi^{\prime 2}+\frac{2i\tilde{m}}{\sqrt{l}}\varphi'\right)\left(1+\mathcal{K}_{j,1}+\mathcal{K}_{j,2}\right).
\end{split}\end{equation}
The integrations over $\varphi, \varphi'$ and $\theta$ are Gaussian and can be performed straightforwardly to give an expansion of the form
\begin{equation}\label{eq3_10_3}
\begin{split}
U^1_{l_j,l_j'}\sim &    \sqrt{\frac{\pi a_1 \tau}{2  l}}
\left(\frac{1+\tau}{1-\tau}\right)^{\frac{l}{2a_1}+\frac{(a_1+1)}{4a_1}n_j+\frac{a_2}{4a_1}n_{j+1}+\frac{q_j}{2}+\frac{D-2}{2} }
 \\&\times
 \exp\left(-\frac{l}{a_1\tau}+\frac{\tau}{2a_1l}\left(\frac{a_1+1}{2}n_j+\frac{a_2}{2}n_{j+1}+a_1q_j\right)^2
 -\frac{l\vep}{a_1\tau}-\frac{\tilde{m}^2}{2\tau a_2l}\right) \left(1+\mathcal{M}_{j,1}+\mathcal{M}_{j,2}\right).
\end{split}\end{equation}
$U^2_{l_j',l_{j+1}}$ is obtained from $U^1_{l_j,l_j'}$ by interchanging $n_j$ and $n_{j+1}$. Namely, we have an expansion of the form
\begin{equation*}
\begin{split}
U^2_{l_j',l_{j+1}}\sim &    \sqrt{\frac{\pi a_1 \tau}{2  l}}
\left(\frac{1+\tau}{1-\tau}\right)^{\frac{l}{2a_1}+\frac{a_2}{4a_1}n_{j}+\frac{(a_1+1)}{4a_1}n_{j+1}+\frac{q_j}{2}+\frac{D-2}{2} }
 \\&\times
 \exp\left(-\frac{l}{a_1\tau}+\frac{\tau}{2a_1l}\left(\frac{a_2}{2}n_{j}+\frac{a_1+1}{2}n_{j+1}+a_1q_j\right)^2
 -\frac{l\vep}{a_1\tau}-\frac{\tilde{m}^2}{2\tau a_2l}\right) \left(1+\mathcal{N}_{j,1}+\mathcal{N}_{j,2}\right),
\end{split}\end{equation*}where $\mathcal{N}_{j,1}$ and $\mathcal{N}_{j,2}$ are obtained respectively from $\mathcal{M}_{j,1}$ and $\mathcal{M}_{j,2}$ by interchanging $n_j$ and $n_{j+1}$.

Next, we consider the asymptotic expansion of $T^2_{l_j'}$.   Debye uniform asymptotic expansions of modified Bessel functions state that \cite{3, 4}:

\begin{align*}
&I_{\nu}(\nu z) \sim \frac{1}{\sqrt{2\pi \nu}}\frac{e^{\nu\eta(z)}}{(1+z^2)^{1/4}} \left(1+\frac{u_1(t(z))}{\nu}\right),\\
&K_{\nu}(\nu z) \sim \sqrt{\frac{\pi}{2 \nu}}\frac{e^{-\nu\eta(z)}}{(1+z^2)^{1/4}} \left(1-\frac{u_1(t(z))}{\nu}\right),\\
&-\frac{D-2}{2}I_{\nu}(\nu z)+\nu zI_{\nu}'(\nu z)\sim  \frac{\sqrt{\nu}e^{\nu\eta(z)}(1+z^2)^{1/4}}{\sqrt{2\pi}}\left(1+ \frac{m_1(t (z))}{\nu }\right),\\
&-\frac{D-2}{2}K_{\nu}(\nu z)+\nu zK_{\nu}'(\nu z)\sim- \sqrt{\frac{\pi\nu}{2}}  e^{-\nu\eta(z)}(1+z^2)^{1/4}\left(1- \frac{m_1(t (z))}{\nu }\right),
\end{align*}where
\begin{align*}
u_1(t )=&\frac{t }{8}-\frac{5t^3}{24},\hspace{1cm}m_1(t )=-\frac{(4D-5)t}{8}+\frac{7t^3}{24},\\
t (z)=&\frac{1}{\sqrt{1+z^2}},\hspace{1cm} \eta(z)=\sqrt{1+z^2}+\ln\frac{z}{1+\sqrt{1+z^2}}.
\end{align*}
Hence,
\begin{align*}
T^{2, \text{D}}_{l_j'}\sim & \frac{1}{\pi}e^{2\nu\eta(z)}\left(1+\frac{2u_1(t(z))}{\nu}\right),\\
T^{2, \text{N}}_{l_j'}\sim & -\frac{1}{\pi}e^{2\nu\eta(z)}\left(1+\frac{2m_1(t(z))}{\nu}\right),
\end{align*}
where
\begin{align*}
\nu=l_j'+\frac{D-2}{2},\hspace{1cm} z=\frac{\omega a_2 }{\nu}.
\end{align*}
Hence, we have an expansion of the form
\begin{equation}\label{eq3_10_4}\begin{split}
T_{l_j'}^{2, \text{Y}}\sim & \frac{(-1)^{\alpha_\text{Y}}}{\pi}\left(\frac{1+\tau}{1-\tau}\right)^{-\frac{la_2}{a_1}-\frac{a_2}{2a_1}(n_j+n_{j+1})-q_j-\frac{D-2}{2}}
\exp\left(\frac{2a_2l}{a_1\tau}-\frac{a_2\tau(n_j+n_{j+1})^2}{4a_1l}-\frac{a_1\tau q_j^2}{a_2l}-\frac{\tau q_j(n_j+n_{j+1})}{l}\right)\\&\times\exp\left(\mathcal{O}_{j,1}+\mathcal{O}_{j,2}\right)\left(1+\mathcal{P}^{\text{Y}}\right),
\end{split}\end{equation}
where $\alpha_{\text{D}}=0, \alpha_{\text{N}}=1$,
\begin{align*}
\mathcal{P}^{\text{D}}=&\frac{a_1}{a_2 l}\left(\frac{\tau }{4}-\frac{5\tau^3}{12}\right),\\
\mathcal{P}^{\text{N}}=&\frac{a_1}{a_2 l}\left(-\frac{(4D-5)\tau }{4}+\frac{7\tau^3}{12}\right).
\end{align*}
The summation over $l_j'$ in \eqref{eq3_10_1} can be replaced by summation over $q$, which, to leading contributions to the Casimir interaction energy, can be approximated by an integration over $q$ from $-\infty$ to $\infty$. From \eqref{eq3_10_2}, \eqref{eq3_10_3} and \eqref{eq3_10_4}, we have
\begin{align*}
\sum_{l_j'=\tilde{m}}^{\infty}U^1_{l_j,l_j'}T_{l_j'}^2U^2_{l_j',l_{j+1}}\sim & \left(\frac{1+\tau}{1-\tau}\right)^{l+\frac{1}{2}(n_j+n_{j+1})+\frac{D-2}{2}}
\exp\left(-\frac{2l}{\tau}+\frac{(1+a_1)\tau\left(n_j^2+n_{j+1}^2\right)}{4l}+\frac{(1-a_1)\tau n_jn_{j+1}}{2l}
 -\frac{2l\vep}{a_1\tau}-\frac{\tilde{m}^2}{\tau a_2l}\right) \\&\times (-1)^{\alpha_\text{Y}}\frac{a_1\tau}{2l}\left(1+\mathcal{P}^{\text{Y}}\right)\int_{-\infty}^{\infty} dq_j
\exp\left( -\frac{a_1^2\tau q_j^2}{a_2l} \right) \left(1+\mathcal{Q}_{j,1}+\mathcal{Q}_{j,2}\right),
\end{align*}
where
\begin{equation*}
\begin{split}
\mathcal{Q}_{j,1}=&\mathcal{M}_{j,1}+\mathcal{N}_{j,1}+\mathcal{O}_{j,1},\\
\mathcal{Q}_{j,2}=&\mathcal{M}_{j,1}\mathcal{N}_{j,1}+\mathcal{M}_{j,1}\mathcal{O}_{j,1}+\mathcal{N}_{j,1}\mathcal{O}_{j,1}+\mathcal{M}_{j,2}+\mathcal{N}_{j,2}+\mathcal{O}_{j,2}+
\frac{1}{2}\mathcal{O}_{j,1}^2.
\end{split}
\end{equation*}
The integration over $q$ is straightforward and we obtain an expansion of the form
\begin{align*}
\sum_{l_j'=\tilde{m}}^{\infty}U^1_{l_j,l_j'}T_{l_j'}^2U^2_{l_j',l_{j+1}}\sim & \left(\frac{1+\tau}{1-\tau}\right)^{l+\frac{1}{2}(n_j+n_{j+1})+\frac{D-2}{2}}
\exp\left(-\frac{2l}{\tau}+\frac{(1+a_1)\tau\left(n_j^2+n_{j+1}^2\right)}{4l}+\frac{ a_2\tau n_jn_{j+1}}{2l}
 -\frac{2l\vep}{a_1\tau}-\frac{\tilde{m}^2}{\tau a_2l}\right) \\&\times \frac{(-1)^{\alpha_\text{Y}}}{2}\sqrt{\frac{\pi a_2\tau}{l}}\left(1+\mathcal{P}^{\text{Y}}\right) \left(1+\mathcal{R}_{j,1}+\mathcal{R}_{j,2}\right).
\end{align*}
Next, we consider the asymptotic expansion of $T_{l_j}^1$. Similar to $T_{l_j'}^2$, we have
\begin{equation}\label{eq3_10_4_2}\begin{split}
T_{l_j}^{1, \text{X}}\sim & \frac{(-1)^{\alpha_\text{X}}}{\pi}C^{n_j-n_{j+1}}\left(\frac{1+\tau}{1-\tau}\right)^{-l-\frac{1}{2 }(n_j+n_{j+1}) -\frac{D-2}{2}}
\exp\left(\frac{2 l}{ \tau}-\frac{ \tau(n_j^2+n_{j+1}^2)}{2 l}  \right)\\&\times\exp\left(\mathcal{S}_{j,1}+\mathcal{S}_{j,2}\right)\left(1+\mathcal{T}^{\text{Y}}\right),
\end{split}\end{equation}
where
\begin{align*}
\mathcal{T}^{\text{D}}=&\frac{1}{  l}\left(\frac{\tau }{4}-\frac{5\tau^3}{12}\right),\\
\mathcal{T}^{\text{N}}=&\frac{1}{  l}\left(-\frac{(4D-5)\tau }{4}+\frac{7\tau^3}{12}\right).
\end{align*}
Thence, $M_{l_j,l_{j+1}}$ has an expansion of the form
\begin{align}\label{eq3_10_6}
M_{l_j,l_{j+1}}^{\text{XY}}\sim \frac{(-1)^{\alpha_{\text{X}}+\alpha_{\text{Y}}}}{2}\sqrt{\frac{ a_2\tau}{\pi l}}
\exp\left( -\frac{a_2\tau\left(n_j-n_{j+1}\right)^2}{4l}
 -\frac{2l\vep}{a_1\tau}-\frac{\tilde{m}^2}{\tau a_2l}\right)  \left(1+\mathcal{T}^{\text{X}}+\mathcal{P}^{\text{Y}}\right) \left(1+\mathcal{U}_{j,1}+\mathcal{U}_{j,2}\right),
\end{align}
where
\begin{align*}
\mathcal{U}_{j,1}=&\mathcal{R}_{j,1}+\mathcal{S}_{j,1}\\
\mathcal{U}_{j,2}=&\mathcal{R}_{j,1}\mathcal{S}_{j,1}+\mathcal{R}_{j,2}+\mathcal{S}_{j,2}+\frac{1}{2}\mathcal{S}_{j,1}^2.
\end{align*}X denotes the boundary condition on the first sphere and Y denotes the boundary condition on the second sphere.

Substitute \eqref{eq3_10_6} into \eqref{eq3_6_1}. To obtain the leading contributions to the Casimir interaction energy, we can replace the summation over $\tilde{m}$ and $l_j$, $0\leq j\leq s$ by the corresponding integration over $\tilde{m}$, $l$ and $n_j$, $1\leq j\leq s$. Using
\begin{align*}
\frac{\tilde{m}\left(\tilde{m}+\frac{D-5}{2}\right)!}{\left(\tilde{m}-\frac{D-3}{2}\right)!}=  \tilde{m}^{D-3}-\frac{(D-3)(D-4)(D-5)}{24}\tilde{m}^{D-5}+\ldots.
\end{align*}
we find that
\begin{equation} \begin{split}
E_{\text{Cas}}^{\text{XY}}\sim &-\frac{\hbar c}{\pi(R_1+R_2)}\frac{1}{(D-3)!}\sum_{s=0}^{\infty}\frac{1}{s+1}\frac{(-1)^{(\alpha_{\text{X}}+\alpha_{\text{Y}})(s+1)}}{2^{s+1}}
\frac{a_2^{\frac{s+1}{2}}}{\pi^{\frac{s+1}{2}}}\int_0^{1} \frac{\tau^{\frac{s-3}{2}}d\tau}{ \sqrt{1-\tau^2}}\int_0^{\infty} dl \, l^{-\frac{s-1}{2}} \int_{-\infty}^{\infty}dn_1\ldots\int_{-\infty}^{\infty} dn_s\\
 & \times \int_0^{\infty}d\tilde{m} \left(\tilde{m}^{D-3}-\frac{(D-3)(D-4)(D-5)}{24}\tilde{m}^{D-5}\right) \exp\left( -\sum_{j=0}^{s}\frac{a_2\tau\left(n_j-n_{j+1}\right)^2}{4l}
 -\frac{2l(s+1)\vep}{a_1\tau}-\frac{(s+1)\tilde{m}^2}{\tau a_2l}\right)  \\
 &\times \left(1+\sum_{j=0}^s\mathcal{U}_{j,1}+\sum_{i=0}^{s-1}\sum_{j=i+1}^s\mathcal{U}_{i,1}\mathcal{U}_{j,1}+\sum_{j=0}^s\mathcal{U}_{j,2}+(s+1)\mathcal{T}^{\text{X}}+(s+1)\mathcal{P}^{\text{Y}}\right). \end{split}
\end{equation}Upon integration with respect to $n_j$, $1\leq j\leq s$, the term $\sum_{j=0}^s\mathcal{U}_{j,1}$ of order $\sqrt{\vep}$ does not give any contribution since it is odd in one of the $n_j$'s, and we have an expansion of the form
\begin{equation} \begin{split}
E_{\text{Cas}}^{\text{XY}}\sim &-\frac{\hbar c\sqrt{a_2}}{2\pi^{\frac{3}{2}} a_1(R_1+R_2)(D-3)!}\sum_{s=0}^{\infty}\frac{(-1)^{(\alpha_{\text{X}}+\alpha_{\text{Y}})(s+1)}}{\left(s+1\right)^{\frac{3}{2}}}
 \int_0^{1} \frac{d\tau}{\tau^{\frac{3}{2}} \sqrt{1-\tau^2}}\int_0^{\infty} dl \, l^{\frac{1}{2}} \\
 & \times \int_0^{\infty}d\tilde{m} \left(\tilde{m}^{D-3}-\frac{(D-3)(D-4)(D-5)}{24}\tilde{m}^{D-5}\right) \exp\left(
 -\frac{2l(s+1)\vep}{a_1\tau}-\frac{(s+1)\tilde{m}^2}{\tau a_2l}\right)  \\
 &\times \left(1+\mathcal{V}+(s+1)\mathcal{T}^{\text{X}}+(s+1)\mathcal{P}^{\text{Y}}\right). \end{split}
\end{equation}
From this, we find that the leading term of the Casimir interaction energy is
\begin{align*}
E_{\text{Cas}}^{0, \text{XY}}= &-\frac{\hbar c\sqrt{a_2}}{2\pi^{\frac{3}{2}} a_1(R_1+R_2)(D-3)!}\sum_{s=0}^{\infty}\frac{(-1)^{(\alpha_{\text{X}}+\alpha_{\text{Y}})(s+1)}}{\left(s+1\right)^{\frac{3}{2}}}
 \int_0^{1} \frac{d\tau}{\tau^{\frac{3}{2}} \sqrt{1-\tau^2}}\int_0^{\infty} dl \, l^{\frac{1}{2}} \\
 & \times \int_0^{\infty}d\tilde{m} \,\tilde{m}^{D-3} \exp\left(
 -\frac{2l(s+1)\vep}{a_1\tau}-\frac{(s+1)\tilde{m}^2}{\tau a_2l}\right) \\
=& -\frac{\hbar c a_2^{\frac{D-1}{2}}}{4\pi^{\frac{3}{2}}a_1(R_1+R_2) }\frac{\Gamma\left(\frac{D-2}{2}\right)}{\Gamma(D-2)}\sum_{s=0}^{\infty}\frac{(-1)^{(\alpha_{\text{X}}+\alpha_{\text{Y}})(s+1)}}{\left(s+1\right)^{\frac{D+1}{2}}}
 \int_0^{1} \frac{d\tau\,\tau^{\frac{D-5}{2}} }{\sqrt{1-\tau^2}}\int_0^{\infty} dl \, l^{\frac{D-1}{2}} \exp\left(
 -\frac{2l(s+1)\vep}{a_1\tau} \right) \\
 =& -\frac{\hbar c (a_1a_2)^{\frac{D-1}{2}}}{2^{\frac{3D-1}{2}}\pi (R_1+R_2)\vep^{\frac{D+1}{2}}}\frac{\Gamma\left(\frac{D+1}{2}\right)}{\Gamma\left(\frac{D-1}{2}\right)}\sum_{s=0}^{\infty}\frac{(-1)^{(\alpha_{\text{X}}+\alpha_{\text{Y}})(s+1)}}{\left(s+1\right)^{ D+1 }}
 \int_0^{1} \frac{d\tau\,\tau^{D-2} }{\sqrt{1-\tau^2}}\\
 =& -\frac{\hbar c (a_1a_2)^{\frac{D-1}{2}}}{2^{\frac{3D+1}{2}}\sqrt{\pi}(R_1+R_2) \vep^{\frac{D+1}{2}}}\frac{\Gamma\left(\frac{D+1}{2}\right)}{\Gamma\left(\frac{D}{2}\right)}\sum_{s=0}^{\infty}\frac{(-1)^{(\alpha_{\text{X}}+\alpha_{\text{Y}})(s+1)}}{\left(s+1\right)^{ D+1 }}.
\end{align*}Namely,
\begin{equation}\label{eq3_10_7}\begin{split}
E_{\text{Cas}}^{0,\text{DD/NN}}\sim &-\frac{\Gamma\left(\frac{D+1}{2}\right)\zeta(D+1)}{2^{\frac{3D+1}{2}}\sqrt{\pi}\Gamma\left(\frac{D}{2}\right)}\left(\frac{R_1R_2}{R_1+R_2}\right)^{\frac{D-1}{2}} \frac{\hbar c}{d^{\frac{D+1}{2}}},\\
E_{\text{Cas}}^{0, \text{DN/ND}}\sim &(1-2^{-D})\frac{\Gamma\left(\frac{D+1}{2}\right)\zeta(D+1)}{2^{\frac{3D+1}{2}}\sqrt{\pi}\Gamma\left(\frac{D}{2}\right)}\left(\frac{R_1R_2}{R_1+R_2}\right)^{\frac{D-1}{2}} \frac{\hbar c}{d^{\frac{D+1}{2}}},
\end{split}\end{equation}which agree with the proximity force approximation \eqref{eq12_5_10}.

The next-to-leading order term of the Casimir interaction energy $E_{\text{Cas}}^{1, \text{XY}}$ can be written as a sum of two terms:
\begin{align*}
E_{\text{Cas}}^{1, \text{XY}}=E_{\text{Cas}}^{1a, \text{XY}}+E_{\text{Cas}}^{1b, \text{XY}}.
\end{align*}
$E_{\text{Cas}}^{1a, \text{XY}}$ vanishes if $D=3, 4$ or 5. For $D\geq 6$,
\begin{align*}
&E_{\text{Cas}}^{1a, \text{XY}}\\= & \frac{(D-3)(D-4)(D-5)}{48}\frac{\hbar c\sqrt{a_2}}{ \pi^{\frac{3}{2}} a_1(R_1+R_2)(D-3)!}\sum_{s=0}^{\infty}\frac{(-1)^{(\alpha_{\text{X}}+\alpha_{\text{Y}})(s+1)}}{\left(s+1\right)^{\frac{3}{2}}}
 \int_0^{1} \frac{d\tau}{\tau^{\frac{3}{2}} \sqrt{1-\tau^2}}\int_0^{\infty} dl \, l^{\frac{1}{2}} \\
 & \times \int_0^{\infty}d\tilde{m} \,\tilde{m}^{D-5} \exp\left(
 -\frac{2l(s+1)\vep}{a_1\tau}-\frac{(s+1)\tilde{m}^2}{\tau a_2l}\right) \\
=& \frac{(D-3)(D-4)(D-5)}{96}\frac{\hbar c a_2^{\frac{D-3}{2}}}{ \pi^{\frac{3}{2}}a_1(R_1+R_2) }\frac{\Gamma\left(\frac{D-4}{2}\right)}{\Gamma(D-2)}\sum_{s=0}^{\infty}\frac{(-1)^{(\alpha_{\text{X}}+\alpha_{\text{Y}})(s+1)}}{\left(s+1\right)^{\frac{D-1}{2}}}
 \int_0^{1} \frac{d\tau\,\tau^{\frac{D-7}{2}} }{\sqrt{1-\tau^2}}\int_0^{\infty} dl \, l^{\frac{D-3}{2}} \exp\left(
 -\frac{2l(s+1)\vep}{a_1\tau} \right) \\
 =& \frac{(D-3)(D-4)(D-5)}{6}\frac{\hbar c (a_1a_2)^{\frac{D-3}{2}}}{2^{\frac{3D+1}{2}}\pi (R_1+R_2)\vep^{\frac{D-1}{2}}}\frac{\Gamma\left(\frac{D-4}{2}\right)}{\Gamma\left(\frac{D-2}{2}\right)}\sum_{s=0}^{\infty}\frac{(-1)^{(\alpha_{\text{X}}+\alpha_{\text{Y}})(s+1)}}{\left(s+1\right)^{ D-1 }}
 \int_0^{1} \frac{d\tau\,\tau^{D-4} }{\sqrt{1-\tau^2}}\\
 =& \frac{(D-3)(D-4)(D-5)}{6}\frac{\hbar c (a_1a_2)^{\frac{D-3}{2}}}{2^{\frac{3D+1}{2}}\sqrt{\pi}(R_1+R_2) \vep^{\frac{D-1}{2}}}\frac{\Gamma\left(\frac{D-3}{2}\right)}{(D-4)\Gamma\left(\frac{D-2}{2}\right)}\sum_{s=0}^{\infty}\frac{(-1)^{(\alpha_{\text{X}}+\alpha_{\text{Y}})(s+1)}}{\left(s+1\right)^{ D-1 }}.
\end{align*}
Namely,
\begin{equation*}
\begin{split}
E_{\text{Cas}}^{1a, \text{DD/NN}}=& \left(1-\delta_{D,3}\right)\left(1-\delta_{D,4}\right)\frac{ (D-5)}{3}\frac{\Gamma\left(\frac{D-1}{2}\right)\zeta(D-1)}{2^{\frac{3D+1}{2}}\sqrt{\pi}\Gamma\left(\frac{D-2}{2}\right)}\left(\frac{R_1R_2}{R_1+R_2}\right)^{\frac{D-3}{2}} \frac{\hbar c}{d^{\frac{D-1}{2}}}\\=&-E_{\text{Cas}}^{0, \text{DD/NN}}
\left(1-\delta_{D,3}\right)\left(1-\delta_{D,4}\right)\frac{ (D-5)(D-2)}{3(D-1)}\frac{\zeta(D-1)}{\zeta(D+1)}\left(\frac{d}{R_1}+\frac{d}{R_2}\right),\\
E_{\text{Cas}}^{1a, \text{DN/ND}}=& \left(1-\delta_{D,3}\right)\left(1-\delta_{D,4}\right)\left(1-2^{-D+2}\right)\frac{ (D-5)}{3}\frac{\Gamma\left(\frac{D-1}{2}\right)\zeta(D-1)}{2^{\frac{3D+1}{2}}\sqrt{\pi}\Gamma\left(\frac{D-2}{2}\right)}\left(\frac{R_1R_2}{R_1+R_2}\right)^{\frac{D-3}{2}} \frac{\hbar c}{d^{\frac{D-1}{2}}}\\=&-E_{\text{Cas}}^{0, \text{DD/NN}}
\left(1-\delta_{D,3}\right)\left(1-\delta_{D,4}\right)\frac{ (D-5)(D-2)}{3(D-1)}\frac{\zeta(D-1)}{\zeta(D+1)}\frac{2^D-4}{2^D-1}\left(\frac{d}{R_1}+\frac{d}{R_2}\right).
\end{split}
\end{equation*}
The second contribution to the next-to-leading order term is given by
\begin{equation} \begin{split}
E_{\text{Cas}}^{1b,\text{XY}}= &-\frac{\hbar c\sqrt{a_2}}{2\pi^{\frac{3}{2}} a_1(R_1+R_2)(D-3)!}\sum_{s=0}^{\infty}\frac{(-1)^{(\alpha_{\text{X}}+\alpha_{\text{Y}})(s+1)}}{\left(s+1\right)^{\frac{3}{2}}}
 \int_0^{1} \frac{d\tau}{\tau^{\frac{3}{2}} \sqrt{1-\tau^2}}\int_0^{\infty} dl \, l^{\frac{1}{2}} \\
 & \times \int_0^{\infty}d\tilde{m} \tilde{m}^{D-3} \exp\left(
 -\frac{2l(s+1)\vep}{a_1\tau}-\frac{(s+1)\tilde{m}^2}{\tau a_2l}\right)   \left(\mathcal{V}+(s+1)\mathcal{T}^{\text{X}}+(s+1)\mathcal{P}^{\text{Y}}\right)\\
 =&-\frac{\hbar c a_2^{\frac{D-1}{2}}}{4\pi^{\frac{3}{2}} a_1 (R_1+R_2) }\frac{\Gamma\left(\frac{D-2}{2}\right)}{\Gamma(D-2)}\sum_{s=0}^{\infty}\frac{(-1)^{(\alpha_{\text{X}}+\alpha_{\text{Y}})(s+1)}}{\left(s+1\right)^{\frac{D+1}{2}}}
 \int_0^{1} \frac{d\tau\, \tau^{\frac{D-5}{2}}}{ \sqrt{1-\tau^2}}\int_0^{\infty} dl \, l^{\frac{D-1}{2}} \exp\left(
 -\frac{2l(s+1)\vep}{a_1\tau} \right)   \\
 & \times  \left(\mathcal{W}+(s+1)\mathcal{T}^{\text{X}}+(s+1)\mathcal{P}^{\text{Y}}\right)\\
=& -\frac{\hbar c (a_1a_2)^{\frac{D-1}{2}}}{2^{\frac{3D-1}{2}}\pi (R_1+R_2)\vep^{\frac{D+1}{2}}}\frac{\Gamma\left(\frac{D+1}{2}\right)}{\Gamma\left(\frac{D-1}{2}\right)}\sum_{s=0}^{\infty}\frac{(-1)^{(\alpha_{\text{X}}+\alpha_{\text{Y}})(s+1)}}{\left(s+1\right)^{ D+1 }}
 \int_0^{1} \frac{d\tau\,\tau^{D-2} }{\sqrt{1-\tau^2}}\mathcal{Z}^{\text{XY}}.\end{split}
\end{equation}
Upon integration with respect to $\tau$, we find that
\begin{equation} \begin{split}
E_{\text{Cas}}^{1b,\text{DD}}=& -\frac{\Gamma\left(\frac{D+1}{2}\right)\zeta(D+1)}{2^{\frac{3D+1}{2}}\sqrt{\pi}\Gamma\left(\frac{D}{2}\right)}\left(\frac{R_1R_2}{R_1+R_2}\right)^{\frac{D-1}{2}} \frac{\hbar c}{d^{\frac{D+1}{2}}} \left(\left[-\frac{(D+1)}{4}\frac{d}{R_1+R_2}+\frac{D+1}{12}\left(\frac{d}{R_1}+\frac{d}{R_2}\right)\right]
\sum_{s=0}^{\infty}\frac{1}{\left(s+1\right)^{ D+1 }}\right.\\&\left.\hspace{4cm}+\left[\left(1-\delta_{D,3}\right)\frac{(D-2)(D-5)}{3(D-1)}-\frac{(D-2)(D-3)}{3D}\right]\left(\frac{d}{R_1}+\frac{d}{R_2}\right)\sum_{s=0}^{\infty}\frac{1}{\left(s+1\right)^{ D-1 }}\right)\\
=&E_{\text{Cas}}^{0,\text{DD}}\left( -\frac{(D+1)}{4}\frac{d}{R_1+R_2}+\frac{D+1}{12}\left(\frac{d}{R_1}+\frac{d}{R_2}\right)
 \right.\\&\left.\hspace{4cm}+\left[\left(1-\delta_{D,3}\right)\frac{(D-2)(D-5)}{3(D-1)}-\frac{(D-2)(D-3)}{3D}\right]\left(\frac{d}{R_1}+\frac{d}{R_2}\right)\frac{\zeta(D-1)}{\zeta(D+1)}\right),\end{split}
\end{equation}\begin{equation} \begin{split}
 E_{\text{Cas}}^{1b,\text{NN}}=& -\frac{\Gamma\left(\frac{D+1}{2}\right)\zeta(D+1)}{2^{\frac{3D+1}{2}}\sqrt{\pi}\Gamma\left(\frac{D}{2}\right)}\left(\frac{R_1R_2}{R_1+R_2}\right)^{\frac{D-1}{2}} \frac{\hbar c}{d^{\frac{D+1}{2}}} \left(\left[-\frac{(D+1)}{4}\frac{d}{R_1+R_2}+\frac{D+1}{12}\left(\frac{d}{R_1}+\frac{d}{R_2}\right)\right]
\sum_{s=0}^{\infty}\frac{1}{\left(s+1\right)^{ D+1 }}\right.\\&\left.\hspace{4cm}+\left[\left(1-\delta_{D,3}\right)\frac{(D-2)(D-5)}{3(D-1)}-\frac{D^2+7D-6}{3D}\right]\left(\frac{d}{R_1}+\frac{d}{R_2}\right)\sum_{s=0}^{\infty}\frac{1}{\left(s+1\right)^{ D-1 }}\right)\\
=&E_{\text{Cas}}^{0,\text{NN}}\left( -\frac{(D+1)}{4}\frac{d}{R_1+R_2}+\frac{D+1}{12}\left(\frac{d}{R_1}+\frac{d}{R_2}\right)
\right.\\&\left.\hspace{4cm}+\left[\left(1-\delta_{D,3}\right)\frac{(D-2)(D-5)}{3(D-1)}-\frac{D^2+7D-6}{3D}\right]\left(\frac{d}{R_1}+\frac{d}{R_2}\right)\frac{\zeta(D-1)}{\zeta(D+1)}\right),\\
\end{split}
\end{equation}
\begin{equation} \begin{split}
 E_{\text{Cas}}^{1b,\text{DN}}=& -\frac{\Gamma\left(\frac{D+1}{2}\right)\zeta(D+1)}{2^{\frac{3D+1}{2}}\sqrt{\pi}\Gamma\left(\frac{D}{2}\right)}\left(\frac{R_1R_2}{R_1+R_2}\right)^{\frac{D-1}{2}} \frac{\hbar c}{d^{\frac{D+1}{2}}} \left(\left[-\frac{(D+1)}{4}\frac{d}{R_1+R_2}+\frac{D+1}{12}\left(\frac{d}{R_1}+\frac{d}{R_2}\right)\right]
\sum_{s=0}^{\infty}\frac{(-1)^{s+1}}{\left(s+1\right)^{ D+1 }}\right.\\&\left.\hspace{1cm}+\left[\left(1-\delta_{D,3}\right)\frac{(D-2)(D-5)}{3(D-1)}-\frac{(D-2)(D-3)}{3D}\frac{d}{R_1}-\frac{D^2+7D-6}{3D}\frac{d}{R_2}\right]\sum_{s=0}^{\infty}\frac{(-1)^{s+1}}{\left(s+1\right)^{ D-1 }}\right)\\
=&E_{\text{Cas}}^{0,\text{DN}}\left( -\frac{(D+1)}{4}\frac{d}{R_1+R_2}+\frac{D+1}{12}\left(\frac{d}{R_1}+\frac{d}{R_2}\right)
\right.\\&\left.\hspace{1cm}+\left[\left(1-\delta_{D,3}\right)\frac{(D-2)(D-5)}{3(D-1)}-\frac{(D-2)(D-3)}{3D}\frac{d}{R_1}-\frac{D^2+7D-6}{3D}\frac{d}{R_2}\right]\frac{2^D-4}{2^D-1}\frac{\zeta(D-1)}{\zeta(D+1)}\right),\\
\end{split}
\end{equation}
\begin{equation} \begin{split}
 E_{\text{Cas}}^{1b,\text{ND}}=& -\frac{\Gamma\left(\frac{D+1}{2}\right)\zeta(D+1)}{2^{\frac{3D+1}{2}}\sqrt{\pi}\Gamma\left(\frac{D}{2}\right)}\left(\frac{R_1R_2}{R_1+R_2}\right)^{\frac{D-1}{2}} \frac{\hbar c}{d^{\frac{D+1}{2}}} \left(\left[-\frac{(D+1)}{4}\frac{d}{R_1+R_2}+\frac{D+1}{12}\left(\frac{d}{R_1}+\frac{d}{R_2}\right)\right]
\sum_{s=0}^{\infty}\frac{(-1)^{s+1}}{\left(s+1\right)^{ D+1 }}\right.\\&\left.\hspace{1cm}+\left[\left(1-\delta_{D,3}\right)\frac{(D-2)(D-5)}{3(D-1)}-\frac{D^2+7D-6}{3D}\frac{d}{R_1}-\frac{(D-2)(D-3)}{3D}\frac{d}{R_2}\right]\sum_{s=0}^{\infty}\frac{(-1)^{s+1}}{\left(s+1\right)^{ D-1 }}\right)\\
=&E_{\text{Cas}}^{0,\text{DN}}\left( -\frac{(D+1)}{4}\frac{d}{R_1+R_2}+\frac{D+1}{12}\left(\frac{d}{R_1}+\frac{d}{R_2}\right)
\right.\\&\left.\hspace{1cm}+\left[\left(1-\delta_{D,3}\right)\frac{(D-2)(D-5)}{3(D-1)}-\frac{D^2+7D-6}{3D}\frac{d}{R_1}-\frac{(D-2)(D-3)}{3D}\frac{d}{R_2}\right]\frac{2^D-4}{2^D-1}\frac{\zeta(D-1)}{\zeta(D+1)}\right).
\end{split}
\end{equation}

Adding together the leading and next-to-leading order term, we find that if $D=3$,
\begin{equation}
\begin{split}
E_{\text{Cas}}^{\text{DD}}=&E_{\text{Cas}}^{0, \text{DD}}\left(1-\frac{d}{R_1+R_2}+\frac{1}{3}\left(\frac{d}{R_1}+\frac{d}{R_2}\right)+\ldots \right),\\
E_{\text{Cas}}^{\text{NN}}=&E_{\text{Cas}}^{0, \text{NN}}\left(1-\frac{d}{R_1+R_2}+\left(\frac{1}{3}-\frac{40}{\pi^2}\right)\left(\frac{d}{R_1}+\frac{d}{R_2}\right)+\ldots \right),\\
E_{\text{Cas}}^{\text{DN}}=&E_{\text{Cas}}^{0, \text{DN}}\left(1-\frac{d}{R_1+R_2}+\frac{1}{3}\frac{d}{R_1}+\left(\frac{1}{3}-\frac{160}{7\pi^2}\right) \frac{d}{R_2} +\ldots \right),\\
E_{\text{Cas}}^{\text{ND}}=&E_{\text{Cas}}^{0, \text{ND}}\left(1-\frac{d}{R_1+R_2}+\left(\frac{1}{3}-\frac{160}{7\pi^2}\right) \frac{d}{R_1}+\frac{1}{3}\frac{d}{R_2} +\ldots \right),
\end{split}
\end{equation}which agree with the results we obtain in \cite{5}.

If $D=4$,
\begin{equation}
\begin{split}
E_{\text{Cas}}^{\text{DD}}=&E_{\text{Cas}}^{0, \text{DD}}\left(1-\frac{5}{4}\frac{d}{R_1+R_2}+\left(\frac{5}{12}-\frac{7}{18}\frac{\zeta(3)}{\zeta(5)}\right)\left(\frac{d}{R_1}+\frac{d}{R_2}\right)+\ldots \right),\\
E_{\text{Cas}}^{\text{NN}}=&E_{\text{Cas}}^{0, \text{NN}}\left(1-\frac{5}{4}\frac{d}{R_1+R_2}+\left(\frac{5}{12}-\frac{61}{18}\frac{\zeta(3)}{\zeta(5)}\right)\left(\frac{d}{R_1}+\frac{d}{R_2}\right)+\ldots \right),\\
E_{\text{Cas}}^{\text{DN}}=&E_{\text{Cas}}^{0, \text{DN}}\left(1-\frac{5}{4}\frac{d}{R_1+R_2}+\left(\frac{5}{12}-\frac{14}{45}\frac{\zeta(3)}{\zeta(5)}\right)\frac{d}{R_1}+\left(\frac{5}{12}-\frac{122}{45}\frac{\zeta(3)}{\zeta(5)}\right) \frac{d}{R_2} +\ldots \right),\\
E_{\text{Cas}}^{\text{ND}}=&E_{\text{Cas}}^{0, \text{ND}}\left(1-\frac{5}{4}\frac{d}{R_1+R_2}+\left(\frac{5}{12}-\frac{122}{45}\frac{\zeta(3)}{\zeta(5)}\right) \frac{d}{R_1}+\left(\frac{5}{12}-\frac{14}{45}\frac{\zeta(3)}{\zeta(5)}\right)\frac{d}{R_2} +\ldots \right).
\end{split}
\end{equation}

If $D\geq 5$, we have
\begin{equation}\label{eq3_13_1}
\begin{split}
E_{\text{Cas}}^{\text{DD}}=&E_{\text{Cas}}^{0, \text{DD}}\left\{1-\frac{D+1}{4}\frac{d}{R_1+R_2}+\left(\frac{D+1}{12}-\frac{(D-2)(D-3)}{3D}\frac{\zeta(D-1)}{\zeta(D+1)}\right)\left(\frac{d}{R_1}+\frac{d}{R_2}\right)+\ldots \right\},\\
E_{\text{Cas}}^{\text{NN}}=&E_{\text{Cas}}^{0, \text{NN}}\left\{1-\frac{D+1}{4}\frac{d}{R_1+R_2}+\left(\frac{D+1}{12}-\frac{D^2+7D-6}{3D}\frac{\zeta(D-1)}{\zeta(D+1)}\right)\left(\frac{d}{R_1}+\frac{d}{R_2}\right)+\ldots \right\},\\
E_{\text{Cas}}^{\text{DN}}=&E_{\text{Cas}}^{0, \text{DN}}\left\{1-\frac{D+1}{4}\frac{d}{R_1+R_2}+\left(\frac{D+1}{12}-\frac{(D-2)(D-3)}{3D}
\frac{2^D-4}{2^D-1}\frac{\zeta(D-1)}{\zeta(D+1)}\right)\frac{d}{R_1}\right.\\&\left.\hspace{3cm}+\left(\frac{D+1}{12}-\frac{D^2+7D-6}{3D}\frac{2^D-4}{2^D-1}\frac{\zeta(D-1)}{\zeta(D+1)}\right) \frac{d}{R_2} +\ldots \right\},\\
E_{\text{Cas}}^{\text{ND}}=&E_{\text{Cas}}^{0, \text{ND}}\left\{1-\frac{D+1}{4}\frac{d}{R_1+R_2}+\left(\frac{D+1}{12}-\frac{D^2+7D-6}{3D}\frac{2^D-4}{2^D-1}\frac{\zeta(D-1)}{\zeta(D+1)}\right) \frac{d}{R_1}\right.\\&\left.\hspace{3cm}+\left(\frac{D+1}{12}-\frac{(D-2)(D-3)}{3D}
\frac{2^D-4}{2^D-1}\frac{\zeta(D-1)}{\zeta(D+1)}\right)\frac{d}{R_2} +\ldots \right\}.
\end{split}
\end{equation}

Notice that if we take the limit the radius of the second sphere is very large, i.e. $R_2\rightarrow\infty$, we recover the results for sphere-plane we obtained in \cite{1}. In fact,  if the Casimir interaction energy for the sphere-plate case is given by
\begin{equation}
\begin{split}
E_{\text{Cas}}^{\text{SP, XY}}=E_{\text{Cas}}^{0, \text{SP, XY}}\left(1+\vartheta^{\text{ XY}}_E\frac{d}{R}+\ldots\right),
\end{split}
\end{equation}where $R$ is the radius of the sphere, X is the boundary condition on the sphere, and Y is the boundary condition on the plate, then the Casimir interaction energy for the sphere-sphere case can be written as
\begin{equation}\begin{split}
E_{\text{Cas}}^{\text{DD}}=&E_{\text{Cas}}^{0, \text{DD}}\left\{1-\frac{D+1}{4}\frac{d}{R_1+R_2}+\vartheta_E^{\text{DD}}\left(\frac{d}{R_1}+\frac{d}{R_2}\right)+\ldots\right\},\\
E_{\text{Cas}}^{\text{NN}}=&E_{\text{Cas}}^{0, \text{DD}}\left\{1-\frac{D+1}{4}\frac{d}{R_1+R_2}+\vartheta_E^{\text{NN}}\left(\frac{d}{R_1}+\frac{d}{R_2}\right)+\ldots\right\},\\
E_{\text{Cas}}^{\text{DN}}=&E_{\text{Cas}}^{0, \text{DD}}\left\{1-\frac{D+1}{4}\frac{d}{R_1+R_2}+\vartheta_E^{\text{DN}} \frac{d}{R_1}+\vartheta_E^{\text{ND}}\frac{d}{R_2} +\ldots\right\},\\
E_{\text{Cas}}^{\text{ND}}=&E_{\text{Cas}}^{0, \text{DD}}\left\{1-\frac{D+1}{4}\frac{d}{R_1+R_2}+\vartheta_E^{\text{ND}} \frac{d}{R_1}+\vartheta_E^{\text{DN}}\frac{d}{R_2} +\ldots\right\}.
\end{split}
\end{equation}
The values of $\vartheta_E^{\text{XY}}$ have been tabulated in \cite{1} for $3\leq D\leq 12$.

Comparing the representation of the Casimir interaction energy we derive in this paper with the one used in \cite{5}, the major difference is that for the elements of the translation matrices $G^1_{\boldsymbol{m},\boldsymbol{m}'}$ and $G^2_{\boldsymbol{m}',\boldsymbol{m}}$, we leave them as   integrals and do not re-express it as combinations of spherical waves with complicated coefficients. When we find the asymptotic behavior of the Casimir interaction energy, it turns out that this does not incur further complications  thanks to the integral formula \eqref{eq3_12_1}, rather than having to rely on integral formulas for those complicated coefficients as in \cite{5}.

\begin{figure}[h]
\epsfxsize=0.5\linewidth \epsffile{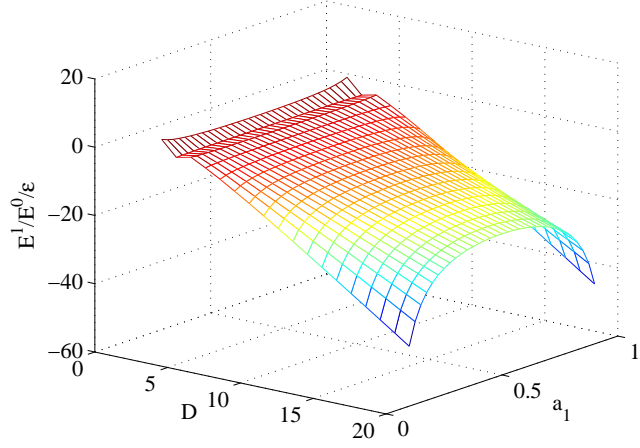} \caption{\label{f1} The dependence of $(R_1+R_2)E^1_{\text{Cas}}/(dE^0_{\text{Cas}})$  on $a_1=R_1/(R_1+R_2)$ and  dimension $D$ for DD boundary conditions.}\end{figure}

\begin{figure}[h]
\epsfxsize=0.5\linewidth \epsffile{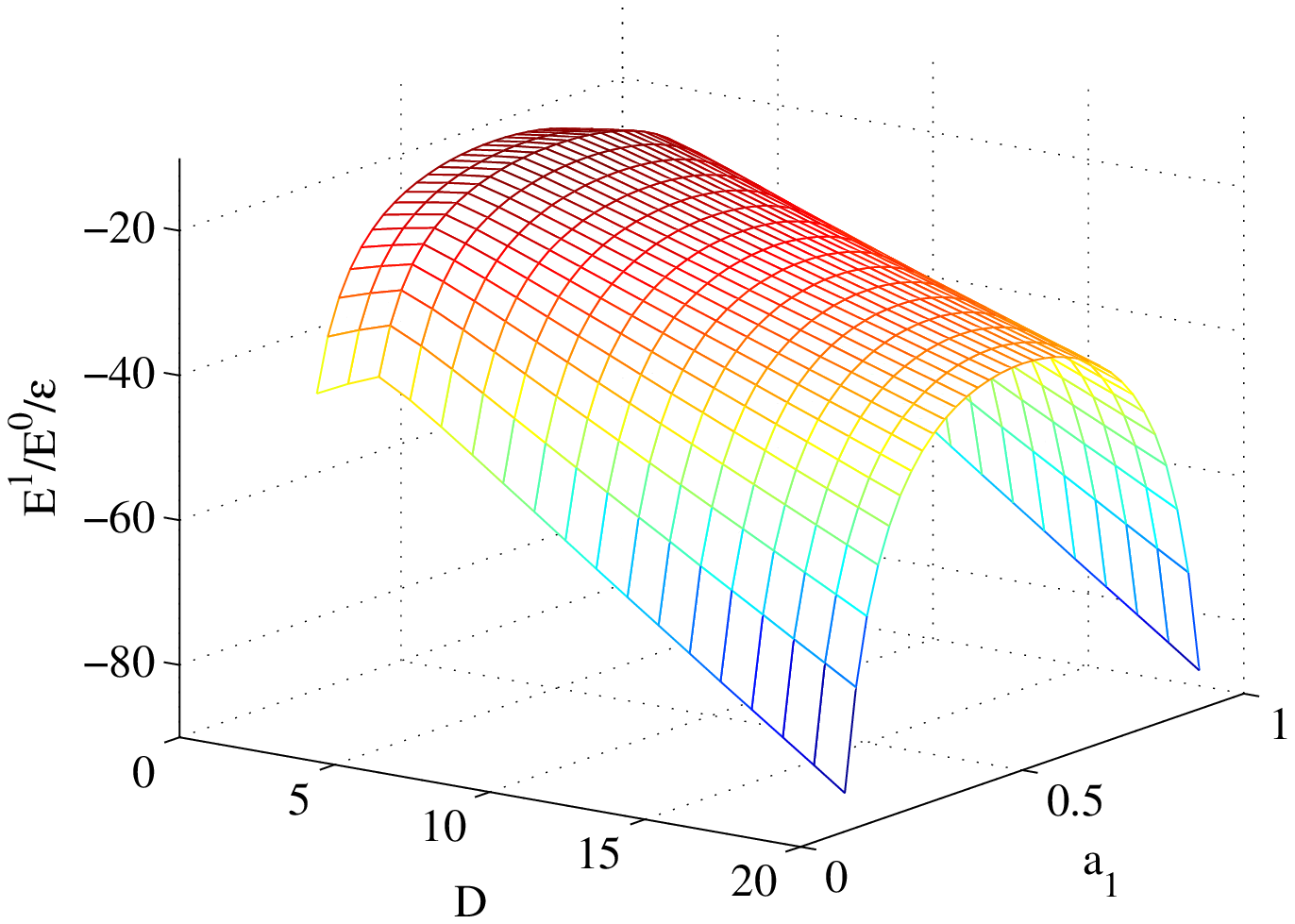} \caption{\label{f2} The dependence of $(R_1+R_2)E^1_{\text{Cas}}/(dE^0_{\text{Cas}})$  on $a_1=R_1/(R_1+R_2)$ and  dimension $D$ for NN boundary conditions.}\end{figure}

\begin{figure}[h]
\epsfxsize=0.5\linewidth \epsffile{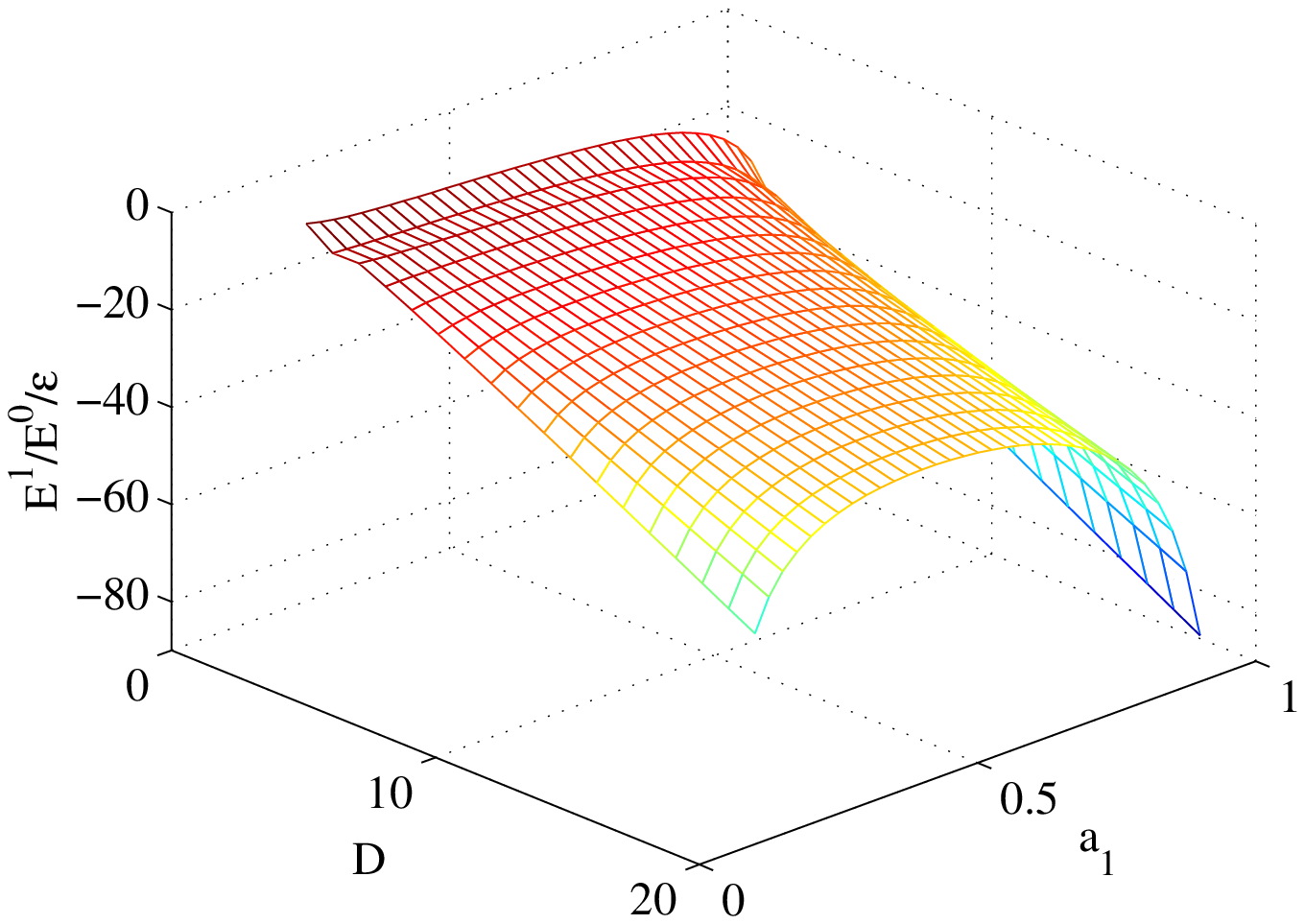} \caption{\label{f3} The dependence of $(R_1+R_2)E^1_{\text{Cas}}/(dE^0_{\text{Cas}})$  on $a_1=R_1/(R_1+R_2)$ and  dimension $D$ for DN boundary conditions.}\end{figure}

\begin{figure}[h]
\epsfxsize=0.5\linewidth \epsffile{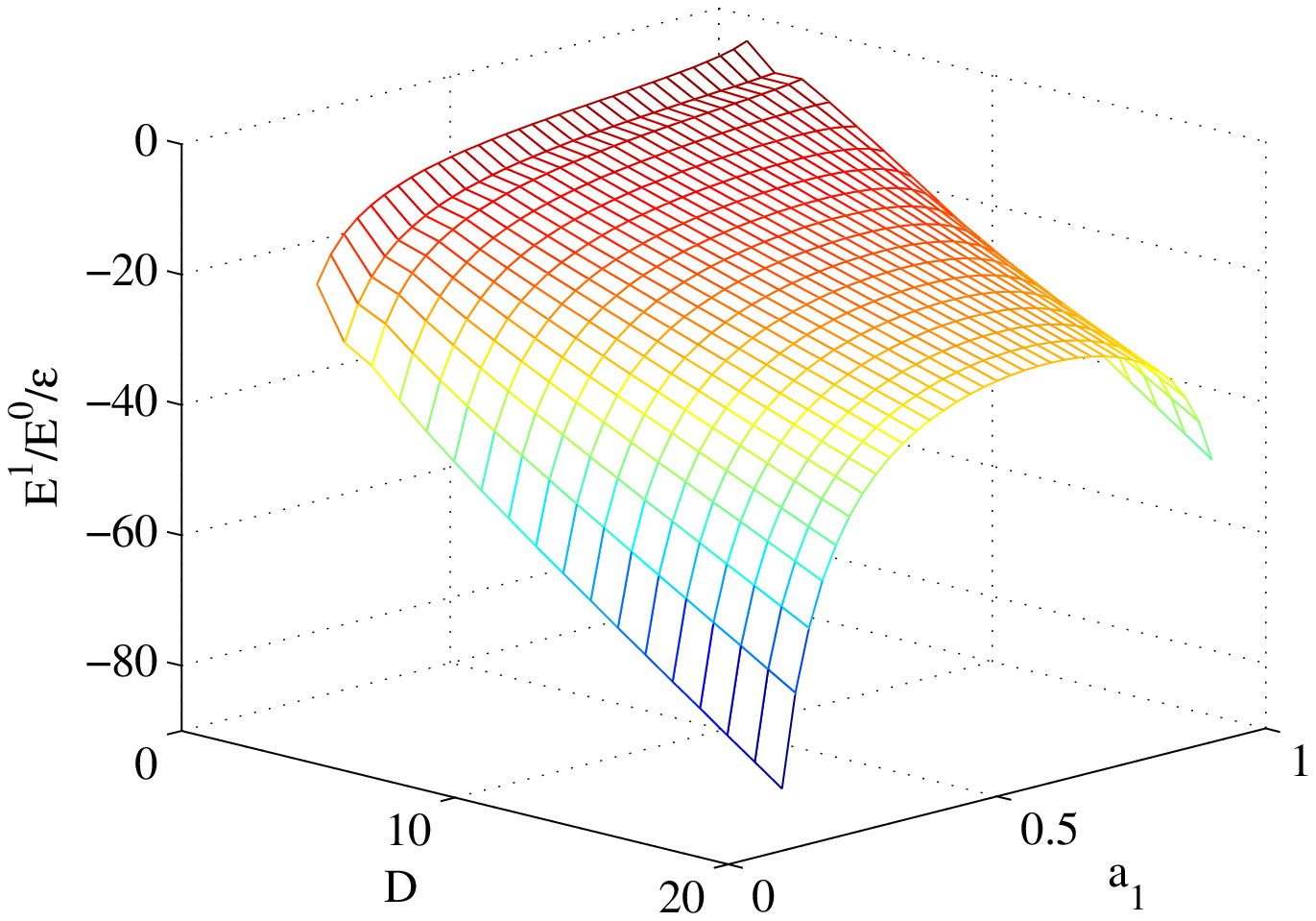} \caption{\label{f4} The dependence of $(R_1+R_2)E^1_{\text{Cas}}/(dE^0_{\text{Cas}})$  on $a_1=R_1/(R_1+R_2)$ and  dimension $D$ for ND boundary conditions.}\end{figure}

Notice that since $a_2=R_2/(R_1+R_2)=1-R_1/(R_1+R_2)$,
\begin{align}\label{eq3_18_1}
\frac{R_1+R_2}{d}\frac{E^1_{\text{Cas}}}{E^0_{\text{Cas}}}= \frac{1}{\vep}\frac{E^1_{\text{Cas}}}{E^0_{\text{Cas}}}
\end{align}is a function of dimension $D$ and $a_1=R_1/(R_1+R_2)$ only. In Figs. \ref{f1}, \ref{f2}, \ref{f3} and \ref{f4}, we plot the dependence of \eqref{eq3_18_1} as a function of  $D$ and $a_1=R_1/(R_1+R_2)$.

We notice that the value of \eqref{eq3_18_1} can become very large for large $D$. In fact, from \eqref{eq3_13_1}, we see that when $D$ is large, the ratio of the next-to-leading order term $E_{\text{Cas}}^1$  to the leading order term $E_{\text{Cas}}^0$  behaves as
\begin{align*}
\frac{E_{\text{Cas}}^1 }{E_{\text{Cas}}^0 }\sim &-\frac{D}{4}\frac{d}{R_1+R_2}+\left(\frac{D}{12}-\frac{D}{3} \right)\left(\frac{d}{R_1}+\frac{d}{R_2}\right)\\
=&-\frac{D}{4}\left(\frac{d}{R_1+R_2}+\frac{d}{R_1}+\frac{d}{R_2}\right)\\
=&-\frac{\vep D}{4}\left(1+\frac{1}{a_1}+\frac{1}{1-a_1}\right),
\end{align*}
which is proportional to $D$. Therefore in higher dimensions, the proximity force approximation to the Casimir interaction energy becomes less accurate, and the contribution of the next-to-leading order term becomes more significant.

\section{Conclusion}

We have derived the TGTG formula for the Casimir interaction energy between two spheres in $(D+1)$-dimensional Minkowski spacetime. The most difficult part in the derivation is the computation of the translation matrices which relate the spherical waves in two coordinate frames differ by a translation. This has not been derived elsewhere and can be considered as a major byproduct of this work. Unlike the three-dimensional case, we do not rewrite the elements of the translation matrices as linear combinations of spherical waves with coefficients expressed in terms of $3j$-symbols. We content with writing them as integrals over Gegenbauer polynomials, which are orthogonal polynomials generalizing Legendre polynomials.

For practical purpose, we explore the strength of the Casimir interaction in the small and large separation regimes.

In the large separation regime, the leading contribution to the Casimir interaction energy comes from a few terms with the lowest wave number(s). It is found that for Dirichlet-Dirichlet, Dirichlet-Neumann and Neumann-Neumann boundary conditions, the leading contributions are of order $L^{-2D+3}$, $L^{-2D+1}$ and $L^{-2D-1}$ respectively, where $L$ is the center-to-center distance of the spheres. Hence in the large separation regime, the Casimir interaction is strongest in the Dirichlet-Dirichlet case, and weakest in the Neumann-Neumann case. It is also observed that the order of the interaction is weaker in higher dimensions.

In the small separation regime, the magnitude of the Casimir interaction is of order $d^{-\frac{D+1}{2}}$, with $d$ the distance between the spheres, which agrees with what predicted by proximity force approximation. One observes that in contrast to large separation, the order of interaction is stronger when the dimension of spacetime is higher. To study the deviation from proximity force approximation, we compute the next-to-leading order term of the Casimir interaction. It is found that the ratio of the next-to-leading term to the leading order term is proportional to $D$, indicating larger corrections in higher dimensions.

In this work, we demonstrate how to compute the Casimir interactions between two spheres in a spacetime with $(D+1)$ dimensions. We only consider scalar field with Dirichlet or Neumann boundary conditions in Minkowski spacetime. Nonetheless, it is easy to see that one can generalize the approach here to any spacetime and any other fields. This also shed some light on how to compute the quantum interaction between two spherical objects in $(D+1)$-dimensional spacetime.

\begin{acknowledgments}\noindent
  This work is supported by the Ministry of Higher Education of Malaysia  under   FRGS grant FRGS/1/2013/ST02/UNIM/02/2.
\end{acknowledgments}

\end{document}